\documentclass[journal]{IEEEtran}
\ifCLASSINFOpdf
\else
\fi
\newtheorem{definition}{Definition}
\newtheorem{lemma}{Lemma}

\usepackage{amsmath}
\usepackage{amssymb}
\usepackage{amsfonts}
\usepackage{graphicx}
\usepackage{algorithm}
\usepackage{algorithmic}
\usepackage{booktabs}
\usepackage{multirow}
\usepackage{threeparttablex}

\usepackage{array}

\usepackage{url}
\usepackage{subfigure}
\usepackage{tabu}

\usepackage{caption}
\begin{document}
\title{Copula Variational LSTM for High-dimensional Cross-market Multivariate Dependence Modeling}
%\title{High-dimensional Cross-market Dependence Modeling by Copula Variational LSTM}


\author{Jia Xu,
        and~Longbing Cao,~\IEEEmembership{Senior~Member,~IEEE}% <-this % stops a space
\IEEEcompsocitemizethanks{\IEEEcompsocthanksitem School of Computer Science, University of Technology Sydney, Australia.\protect\\
E-mail: longbing.cao@uts.edu.au
%\IEEEcompsocthanksitem 
}% <-this % stops an unwanted space
\thanks{Manuscript received April 19, 2005; revised August 26, 2015.}}

\maketitle

\begin{abstract}
We address an important yet challenging problem - modeling high-dimensional dependencies across multivariates such as financial indicators in heterogeneous markets. In reality, a market couples and influences others over time, and the financial variables of a market are also coupled. We make the first attempt to integrate variational sequential neural learning with copula-based dependence modeling to characterize both temporal observable and latent variable-based dependence degrees and structures across non-normal multivariates. Our variational neural network WPVC-VLSTM models variational sequential dependence degrees and structures across multivariate time series by variational long short-term memory networks and regular vine copula. The regular vine copula models non-normal and long-range distributional couplings across multiple dynamic variables. WPVC-VLSTM is verified in terms of both technical significance and portfolio forecasting performance. It outperforms benchmarks including linear models, stochastic volatility models, deep neural networks, and variational recurrent networks in cross-market portfolio forecasting.

%In the increasingly connected world, many systems are more or less coupled with each other in various ways. A typical example is the cross-market portfolio management, where the products of heterogeneous markets are selected and configured for investment. In such cross-market problems, one market is coupled with and influenced by others, and the financial variables of a market are coupled over time. This work makes the first attempt to model both the observation-based and latent dependence degrees and structures of high-dimensional financial variables in cross-market portfolios. It integrates the distribution-based sequential modeling of multivariate time series by variational recurrent neural networks and the regular vine copula-based dependence structures for modeling cross-market dependencies. Our method addresses the needs and gaps of modeling non-normal and long-range distributional interactions across multiple dynamic variables. We verify the model in terms of both technical significance and portfolio investment performance against benchmarks including linear models, stochastic volatility models, deep neural networks, and variational recurrent networks for portfolio forecasting.
\end{abstract}


% \fi
\IEEEpeerreviewmaketitle

\section{Introduction}
\label{sec:introduction}


%\IEEEPARstart{T}{his} demo file is intended to serve as a ``starter file''
%for IEEE Computer Society journal papers produced under \LaTeX\ using
%IEEEtran.cls version 1.8b and later.
%I wish you the best of success.



\textit{High-dimensional cross-multivariate dependencies} are commonly seen in applications with multivariates which couple, co-move or co-influence with each other \cite{RenGLWYC22,Yang23}. Cross-market portfolio management is a perfect example \cite{ang2002asymmetric}, where financial variables of different markets (e.g., equity markets, fixed income markets, foreign exchange markets, and commodity markets) may strongly or weakly couple and influence each other over time. 
%These markets are more or less coupled with each other in terms of their financial indicator interactions. Each data-intensive financial market has its own constituents, processes, and dynamics. 
This drives a new area - \textit{cross-market modeling} (CMM, or cross-market analytics) \cite{Cao20-review,Cao20-techniques,CaoYY21} characterizing the interactions and dynamics both within and between individual markets and their constituents (e.g., participant behaviors, choices and interactions). However, CMM is highly challenging, especially when high-dimensional heterogeneous cross-market financial variables are involved and when they present diverse and non-normal time-series conditions \cite{xu2017copula,Wei2012,cao2011coupled,cao2015deep,CaoC15,CaoDCZ15,SongC12,SongCWWYD12}. We make the first attempt to model both high-dimensional cross-multivariate dependence degrees and structures by integrating variational neural sequential learning with joint probabilistic modeling under non-normal multivariate conditions. We illustrate the design in modeling cross-market dependence degree and structure with complex market dynamics.
%Below, we briefly discuss the challenges in modeling cross-market interactions, the gaps in the related work, and our contributions to effectively integrating statistical dependence modeling, variational autoencoder, and long short-term memory (LSTM) networks for cross-market dependence modeling and portfolio forecasting.

\subsection{Challenges in Cross-multivariate Modeling}
\label{subsec:challenges}

%Financial market analysis has been a lasting research concern with increasingly advanced methods developed from time series regression to deep neural networks-based forecasters over the past decades \cite{Cao20-techniques,Dunis19}. 

The challenges of \textit{high-dimensional cross-multivariate dependence modeling} can be illustrated by CMM. CMM aims to select appropriate financial markets and analyze their couplings and influence on each other \cite{cao2011coupled,Cao20-review,Underwoods07}. CMM goes beyond single financial time series \cite{TranIKG19} and becomes increasingly imperative for regional-to-global financial crisis management, cross-market asset management, portfolio investment, and risk management \cite{Wei2012,xu2017copula}. CMM cannot be handled by conventional studies on time series forecasting \cite{Andersen2009,Gooijer06,TranIKG19} and AI and machine learning \cite{Cao20-techniques} of single market trends and market/price movement. CMM requests further significant efforts on key aspects \cite{Cao20-review,Cao20-techniques} such as: (1) modeling cross-market interactions and influence where there are explicit and implicit inter-financial variables (products) and inter-market coupling relationships \cite{cao2011coupled,cao2015coupling}; (2) characterizing the latent and deep interactions and couplings between financial variables that cannot be represented by time series analysis and shallow machine learners \cite{ang2002asymmetric,bartram2001international,CaoDCZ15}; and (3) transforming learning tasks from standardized and simplified market settings to real market conditions by addressing those less-to-rarely explored characteristics and challenges of financial markets, such as non-IID, nonstationary, volatile, asymmetrical and nonlinear market contexts and characteristics \cite{Cao20-review,patton2004out,AhnLOK09,Wei2012,CaoDCZ15,xu2017copula}.

\begin{figure}[htbp]
\centering
\includegraphics[width=0.48\textwidth]{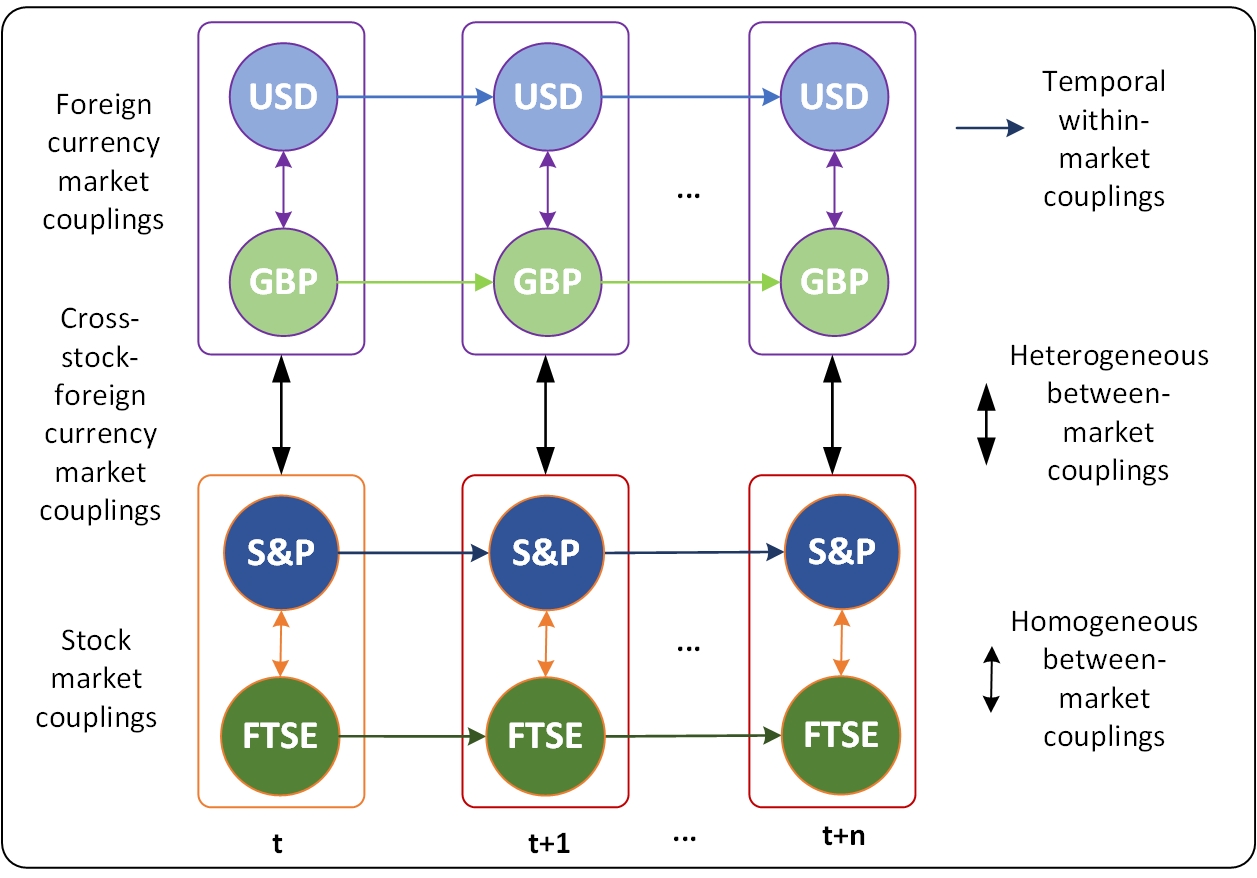}
\caption{Illustration of cross-market interactions: homogeneous and heterogeneous couplings within and between stock and foreign currency markets in the UK and US.}
\label{Coupling1}
\end{figure}

These cross-market couplings and interactions present significant challenges to learning systems. Cross-market couplings may present in the form of explicit observational correlations and dependencies and implicit features and relations. In Fig. \ref{Coupling1}, the temporal movement of the USD exchange rate is coupled with and influenced by three major types of interactions across different markets: (1) \textit{temporal within-market couplings}: the dependence and influence across different time spots within a market, e.g., temporal dependencies between time points $t$ and $t+n$ of the USD exchange rate series; (2) \textit{homogeneous between-market couplings}: the influence between homogeneous  markets, e.g., between the USD and GBP exchange rates; and (3) \textit{heterogeneous between-market couplings}: the interactions and influence between heterogeneous markets, e.g., between the UK stock market index FTSE and US currency USD. In reality, such inter- and intra-market couplings and influences are \textit{temporary} and \textit{heterogeneous} with diverse conditions, distributions and relationships; \textit{dynamic} with self and mutual evolution and cointegration; and \textit{hierarchical} from a financial variable to a market and to these coupled markets. These cross-market couplings are naturally and deeply embedded within and between financial markets. They not only form intrinsic characteristics of individual and collective financial markets but also significantly influence any further derivatives on top of the markets. 
%Therefore, cross-market couplings should be incorporated into estimating the dynamics of financial markets and tasks including portfolio investment, and risk management.

\begin{figure}[h]
  \centering
  \tiny
  \subfigure[FTSE 100]{\label {CrossMartks:2}\includegraphics[width=0.16\textwidth]{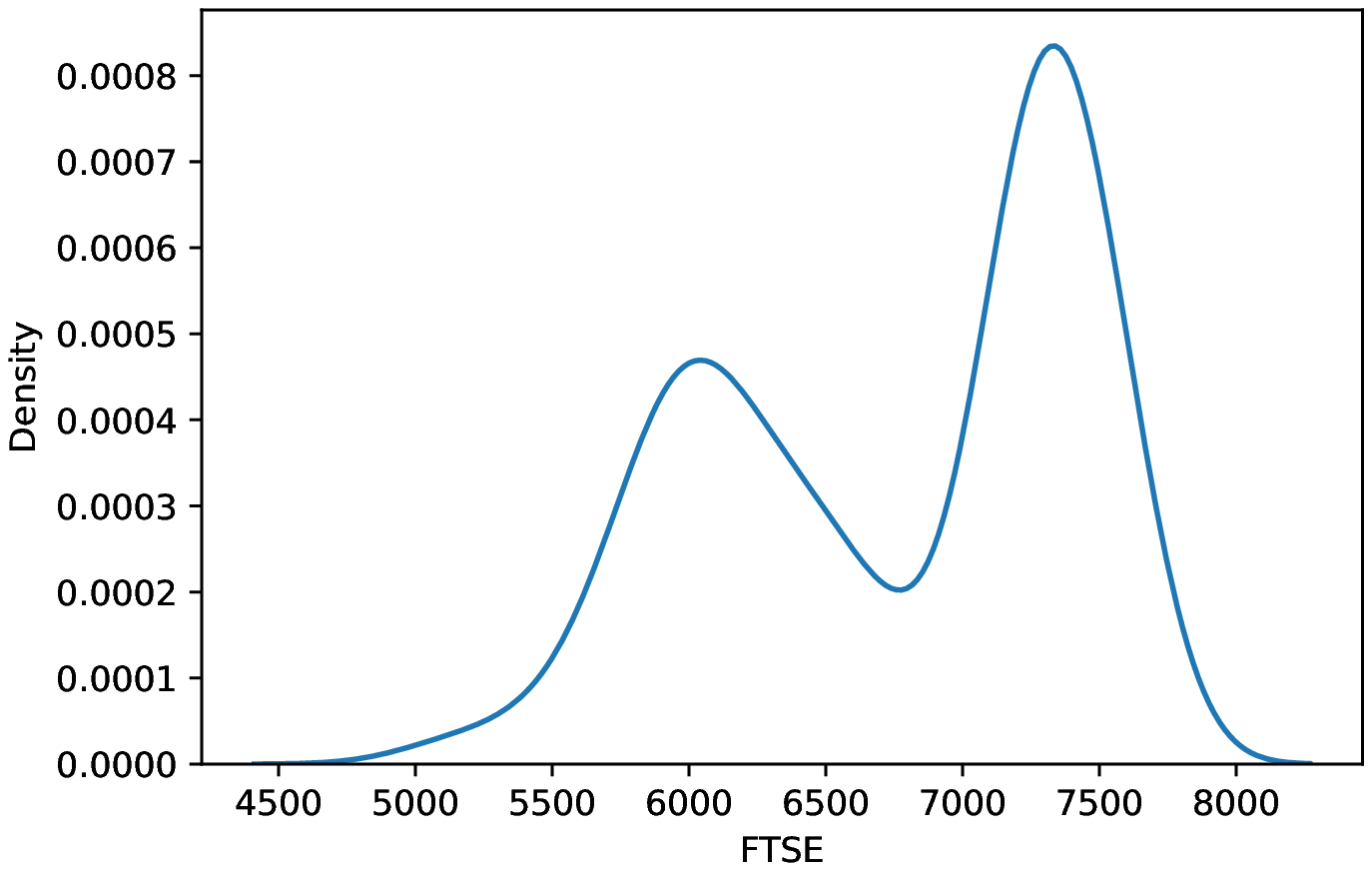}}
  \subfigure[S\&P 500]{\label {CrossMartks:1}\includegraphics[width=0.16\textwidth]{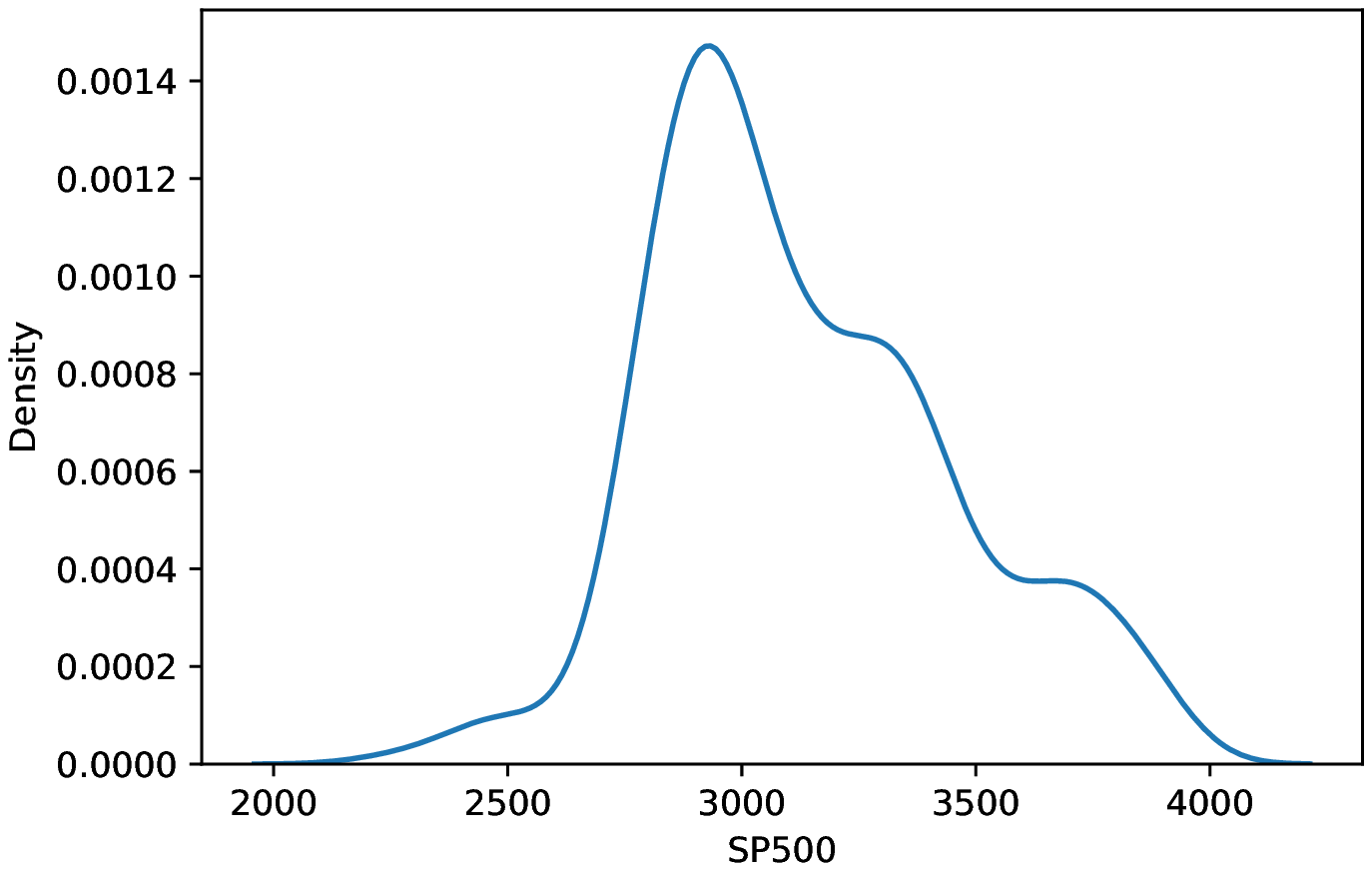}}
  \subfigure[FTSE 100 and S\&P 500]{\label {CrossMartks:3}\includegraphics[width=0.16\textwidth]{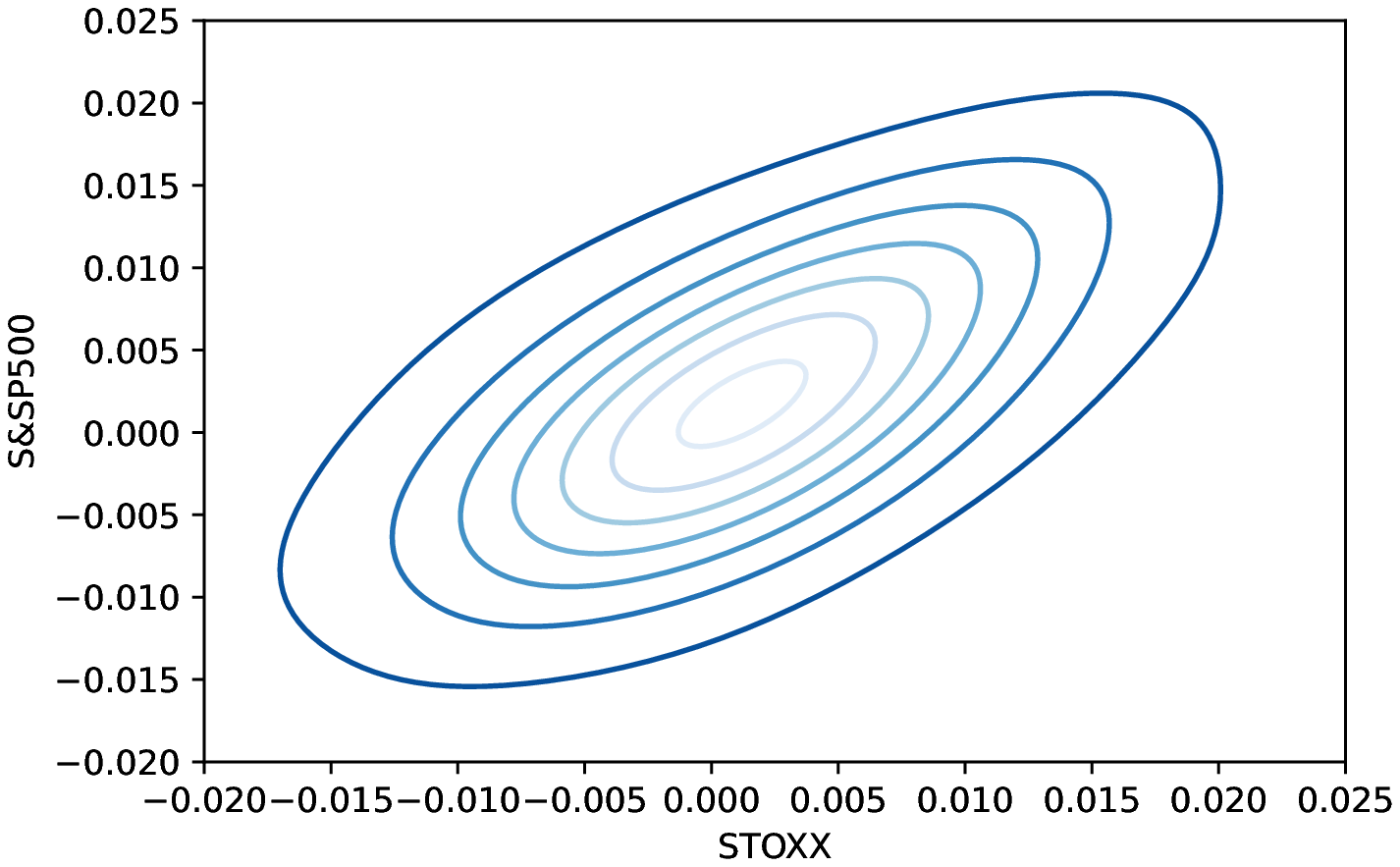}}
 % \\
  %\tiny
  \caption{Illustration of distinct multi-market conditions: heterogeneous distributions and asymmetric couplings between financial markets.}
  \label{CrossMartks}
\end{figure}

However, earlier studies have shown the challenges in modeling such sophisticated cross-market couplings such as for market manipulation, financial crisis contagion, and portfolio management \cite{cao2011coupled,Wei2012,cao2015deep,CaoDCZ15,xu2017copula}. Real market conditions including high frequency, asymmetry, stylist and heterogeneity further complicate the CMM \cite{Cao20-review,Wei2012,xu2017copula,BuczynskiCS21}. 
%Below, we further explore the challenges in cross-market coupling learning from three learning perspectives. 
First, cross-market couplings are associated with various explicit and implicit market factors, likely coupled in complex structures. Accordingly, the first corresponding challenge is to characterize both explicit and implicit couplings in flexible coupling structures without imposing strong assumptions or restrictions on the structures. 
%This is essential for making market coupling learning flexible in characterizing cross-market conditions. 
Second, financial variables and their couplings across markets are non-normal with asymmetrical and nonlinear market characteristics, as shown in Fig. \ref{CrossMartks}. They cannot be well represented by conventional assumptions such as Gaussian, balanced, skewedless, symmetric, stationary and IID (i.e., samples are drawn from the independent and identical distribution) assumptions typically taken in traditional time series analysis, correlation and dependence modeling, and machine learning in finance \cite{boyer1997pitfalls,embrechts2002correlation,ang2002asymmetric,patton2004out,Gooijer06,Andersen2009,Wei2012}. Lastly, multi-markets tend to involve both fundamental factors and exterior information that jointly and lastingly interact with each other and influence the markets over time. Accordingly, modeling cross-market couplings often involves abundant financial and non-financial variables and their interactions, resulting in high-dimensional couplings, which often involve hundreds to thousands of combinations of heterogeneous and coupled input variables. The high-dimensional CMM of many financial time series easily causes the curse of dimensionality. 

For example, the empirical distributions of daily returns of the comprehensive index FTSE100 from the UK and the comprehensive index S\&P500 from the US in Fig. \ref{CrossMartks} are asymmetrical and tend to display much more kurtosis with a pronounced higher peak than allowed under the normality hypothesis. Typically, the return of FTSE100 shows a left-skewed distribution with a fat left tail, while the return of S\&P500 shows a right-skewed distribution with a fat right tail. On the other hand, the correlation between FTSE100 and S\&P500 shown in Fig. \ref{CrossMartks:3} is highly positive with a stronger correlation on the low-bound side, which indicates that these two markets have a weaker correlation in a bull market upturn than in a bear market downturn. 
%This example shows some challenges in modeling cross-market dynamics and couplings. 

\subsection{Gaps in Learning Cross-multivariate Dependencies}

Limited work has been available on learning high-dimensional cross-multivariate dependencies but even less on applying deep learning \cite{RenGLWYC22,Yang23}. Existing studies \cite{Cao20-review,Gooijer06} have focused on financial market analysis by methods including mathematical modeling, AI, and machine learning. They have significant gaps in CMM, which either do not comprehensively address the above challenges or simplify their complexities using much standardized assumptions on market conditions. First, the primary \textit{econometric approaches} \cite{Gooijer06} implement traditional statistical models as demonstrated by the popular auto-regressive moving average (ARMA) and the generalized autoregressive conditional heteroscedasticity (GARCH) families, including those for univiriate and multivariate time series \cite{bauwens2006multivariate}. These approaches apply a deterministic linear function composed of the variables from past observations and their lagged items to formulate the time-varying volatility of financial indicators. The typical linear, stationary and normal assumptions limit their performance and practical applications. Second, \textit{machine learning models} have played a prominent role in recent financial data analysis \cite{Cao20-techniques,SahaGG22}, as typically illustrated by artificial neural networks (ANNs), hidden Markov models (HMMs), and support vector regression (SVR) to forecast financial markets and index/price movements. These shallow learners typically involve standard assumptions to model small-scale and simplified financial markets. For example, HMMs take the first-order relations between variables with all hidden states and observed variables following a Gaussian distribution \cite{cao2011coupled}. When substantial hidden states are introduced to a setting with very high dimensions, on every occasion, it leads to a computational explosion when inferring hidden states from observations. Although ANNs perform well in financial forecasting by more effectively modeling the linear or nonlinear relations between variables, their performance is strongly sensitive to the optimal fitting between input features and network parameters. There could be a risk of overfitting limited observations or there could be a bias to network activation and structures. Shallow ANNs also suffer from being incapable of representing complex factors, deep dependencies, and rich interactions hidden in the data.

\textit{Deep financial modeling} \cite{Cao20-review,CaoYY21} emerges by applying DNNs to financial markets  \cite{Sezeretal19,Ozbayoglu20,Cao20-techniques,WeiC20,ItoSITY20,LiP22}. For example, recurrent neural networks (RNNs) including LSTM and deep state space model, deep factor model, and deep AR model \cite{rangapuram2018deep,salinas2020deepar} process sequential data by many hidden layers and deep hierarchical representations of previous market states transmitted to the next step and layer to deeply model hidden factors, relations and long-range dependencies for financial market forecasting. They require supervised information and large data but are vulnerable to overfitting limited data points and challenged by sophisticated interactions and non-IID characteristics of market dynamics and volatility \cite{cao2011coupled,Cao20-techniques}. DNNs and LSTM are significantly challenged by the above cross-market couplings and complex characteristics \cite{Yang23}.

Alternate to point-based fitting by RNNs, variational autoencoders (VAEs) \cite{kingma2013auto} implement distribution-based probabilistic modeling by involving variational inference (VI) in approximating the posterior over continuous latent variables. Variational RNNs (VRNNs) \cite{chung2015recurrent} further hybridize the merits of VAEs and RNNs to integrate the correlations between latent random variables over neighboring time steps. These VAE networks typically take the mean-field assumption, i.e.,  latent variables are independent for tractability and the posterior is fully factorizable, which may cause the Kullback-Leibler (KL) divergence to vanish during parameter estimation and cannot capture the non-IIDness across variables \cite{Cao2013}. Then, copula functions \cite{Sklar1959} augment the variational inference for more flexible dependence modeling in factorizable distributions \cite{tran2015copula,wang2019neural,BahugunaK20}. They only involve empirical density or standard copula families (e.g., a multivariate Gaussian copula), which cannot characterize the aforementioned hierarchical and heterogeneous cross-market couplings and non-normal market characteristics. Despite being at their early stage, these recent studies represent the state-of-the-art in modeling latent and deep interactions and long-range dependencies in CMM \cite{Cao20-techniques}. 

%In summary, the existing studies do not integratively address the aforementioned major challenges and their derived ones. They typically do not work well for high-dimensional cross-multivariate modeling. This work is motivated to address these gaps.

\subsection{Our Contributions}
A novel general regular vine copula-based variational LSTM network \textit{weighted partial regular vine copula-based variational LSTM} (WPVC-VLSTM) makes the first attempt to jointly model high-dimensional dependencies, long-range dependence structures, non-normal temporal multivariates, and latent features and relations across heterogeneous market variables. WPVC-VLSTM integrates the vine copula-based probabilistic modeling of high-dimensional dependencies and their structures between multiple time series with non-normal characteristics and variational LSTM (VLSTM) for probabilistic recurrent neural learning of long-range dependencies between random variables. The regular vine copula models the dependence structures over modeling hidden variables of VLSTM over time. 

%WPVC-VLSTM represents the first attempt to jointly model probabilistic long-range dependencies over high-dimensional non-normal variables and the dependence structures over modeling parameters. 
WPVC-VLSTM thus combines variational neural learning with probabilistic dependence structure modeling. This work makes the following major contributions: (1) A parameterized regular vine copula with unrestricted abundant variational families of copula better estimates the joint density of posterior distributions. (2) A parametrization technique for vine copula VAE averts the posterior collapse issue. (3) An efficient framework integrates the vine copula VAE with LSTM to capture the diverse dependencies and volatilities between different variables without imposing any Gaussian assumption on the multivariates. WPVC-VLSTM models cross-market couplings across financial markets under complex market conditions. We evaluate its cross-market portfolio forecasting over eight exchange rates and eight comprehensive indices. The experiments show the significant performance of WPVC-VLSTM in comparison with multivariate regressors, convolutional neural networks (CNNs), LSTM, and VLSTM in terms of both technical measures and business impact on portfolio management.

%The rest of the paper is structured as follows. Section 2 summarises the related work. Section 3 briefly presents the background knowledge and foundations for some significant concepts, including copula, VAE, and LSTM. Then the next sections illustrate the WPVC-VLSTM including its model construction and estimation for parameters. In Sections 6 and 7, evaluation metrics for verifying the performance of modeling high-dimensional variables in financial markets are discussed, and case study results are shown. After making discussion on the model in Section 8, we conclude the work in Section 9. 

\section{Related Work}

High-dimensional cross-multivariate dependence modeling involves various related areas including multivariate correlation and dependence modeling, shallow learning of market relations, learning cross-market couplings, and deep sequential modeling and their applications in finance such as CMM \cite{Gooijer06,Cao20-review,Cao20-techniques,CaoYY21}. 

%Financial market analysis has been vigorously discussed in many disciplines including statistics, AI and machine learning in the past decades \cite{Gooijer06,Cao20-review,Cao20-techniques,CaoYY21}. Here, we mainly focus on the work on multivariate time series analysis and shallow and deep machine learning of the relations between multiple financial variables. We also summarize the progress made in CMM.

\subsection{Multivariate Correlation and Dependence Modeling}

Classic statistical models like ARMA and its variants have been widely deployed in  financial time-series prediction \cite{Gooijer06}. To resolve the strong fluctuation clustering issues within financial time series, deterministic linear models (e.g., GARCH and MGARCH and their variants \cite{bauwens2006multivariate}) are introduced to present the residuals in ARMA. For the underlying complex dependencies across financial markets, the most popular solution is to model time-series correlations by ARMA-GARCH models. A classic approach directly applies a joint distribution with a multivariate normal distribution for simplification. Then, inference is conducted by implementing the mean-variance analysis under the Gaussian assumption. However, a large body of evidence demonstrates that it is inappropriate for the Gaussian assumption to capture the dependencies between real-world markets which show significant non-Gaussian characteristics \cite{longin2001extreme}. Further, these joint distribution-based methods only capture dependencies but ignore dependence structures and treat multiple variables as IID \cite{cao2011coupled}. 

As shown in Fig. \ref{CrossMartks} and the relevant literature \cite{Kurowicka2011,Wei2012,Czado2013,xu2017copula}, both  dependence degree and structure should be considered in characterizing cross-market couplings. A number of methods extend the basic correlation analysis, such as in the dynamic conditional correlation (DCC) model \cite{engle2002dynamic}, which calculate the conditional correlations between variables. However, the correlations calculated under different assumptions could be dramatically different. More specifically, correlations conditioned on small movements are weaker than those conditioned on large movements. This is because the level of dependence predicted by a stationary Gaussian process displays dramatic differences at different periods. A volatile period leads to a stronger dependence, while a tranquil period results in a weaker one. Another defect of the DCC family models is that either a very large number of parameters are introduced when no restriction is applied on the dependence structure, which is the co-variance matrix in the model, or the dependence structure is inflexible if restrictions are imposed. 

To address these  issues, copula-based models \cite{demarta2005t} have been shown to be effective in capturing dependencies with complex structures for random variables. Such approaches  inherit the advantages of copulas and are not restricted by linear correlation assumptions, thus they are capable of capturing time-varying dependencies and correlations. In addition to modeling dependencies, vine copula \cite{Joe1997} further provides abundant solutions to enable flexible structures in modeling complex high-dimensional dependence structures. However, due to a lack of efficiency, strong assumptions on vine structures are always implied in most  existing vine-based copula models \cite{Dissmann2013,Wei2012}, imposing hurdles in capturing the aforementioned complex couplings across markets.

\subsection{Shallow Learning of Financial Time-series}

Classic machine learning methods have been widely applied in financial data analysis and modeling financial factor relations \cite{Cao20-techniques,SahaGG22}. Typically, HMMs and their variants capture Markovian processes in finance and represent probabilistic distributions over sequences of financial observations. To resolve the drawbacks of simple HMMs in capturing complex temporal relationships, rich hidden representations are incorporated in some HMM generalizations, such as factorial HMMs \cite{ghahramani1997factorial}, tree structured HMMs \cite{romberg2001bayesian}, and switching state-space models \cite{ghahramani2000variational}. However, when inferring hidden states  from observations, they incur computational intractability when they have richer hidden state representations \cite{ghahramani2001introduction}. 

Another approach is to hybridize time-series models with shallow learners including ANNs, SVM, and SVR. Examples include integrating ANN with GARCH, forming models such as NN-GARCH/ANN-GARCH, NN-APGARCH and LSTM-GARCH/GARCH-LSTM etc models \cite{BILDIRICI20097355,kim2018forecasting}, combining SVM and SVR with AR and GARCH models \cite{chen2010forecasting}, and their variants \cite{StefaniBCHB19}. These methods typically apply one model (e.g., GARCH or MGARCH) to learn time-series as the input to learner such as ANN for forecasting. They do not capture cross-market couplings and cannot compare with the state-of-the-art DNN-based models.

\subsection{Learning Cross-market Couplings}

Modeling cross-market couplings is an attempt to address the aforementioned fundamental challenges such as  coupled group trading behaviors, and multi-market interactions. 
A behavior model \cite{Cao10} captures market or trading activities and their properties. Then, intra- and inter-behavior couplings and their integration are learned to model \textit{coupled group behaviors} \cite{cao2011coupled}; and intra- and inter-market couplings and their integration are learned for cross-market modeling \cite{cao2015deep}. Accordingly, coupled group behaviors in market and cross-market behavior couplings are captured by learning the explicit and implicit couplings within and between multiple financial time series, different stock market factors (e.g., trading price, volume, and actions), and between different financial markets (e.g., equity, commodity, and exchange rate). 

Coupled hidden Markov models (CHMM) are applied to coupled behavior analysis and CMM, including integrated into DNNs \cite{cao2011coupled,SongCWWYD12,SongC12,CaoC15,CaoDCZ15}. Due to the natural limitation of Markovian structures, CHMMs cannot characterize long-range dependencies and complex dependence structures across financial variables and markets, nor are  non-normal conditions captured. 

Another set of studies applies copula dependence modeling \cite{Kurowicka2011} to cross-market coupling learning \cite{Wei2012}. Simple copula functions like Gaussian copula, canonical vine copula, and t copulas \cite{De2010,Dissmann2013,letham2014latent} capture joint dependencies with standard statistical assumptions on market conditions. The recent work further tackles complex market conditions, including the asymmetry, tail distribution and stylist of financial variables \cite{xu2017copula}. Only recently, standard copulas have been combined with DNNs for both neural and probabilistic relational learning \cite{tran2015copula,wang2019neural}. Other methods apply vine copulas to express a broad range of dependencies of posteriors \cite{Wei2012,Dissmann2013,Czado2013,xu2017copula}, which estimate an empirical distribution rather than a joint distribution due to the limitations of vine copula construction. 

\subsection{Deep Sequential Modeling in Finance}

First, DNNs are applied to predict financial time series and macro-economic forecasting \cite{Cao20-techniques}. For example, a coupled temporal deep belief network (CTDBN) \cite{cao2015deep} accommodates the dependencies across financial markets by  a conditional Gaussian restricted Boltzmann machine (RBM) and a coupled conditional RBM in a deep hierarchical structure to capture high-level coupled features. Although CTDBN explores different dependencies, including the temporal couplings within and between financial markets, it is not capable of describing the  asymmetric and nonlinear features of financial variables based on Gaussian assumptions \cite{KhattreeB19}.

RNNs are innate for modeling sequential dependencies in temporal financial variables and markets \cite{Cao20-techniques,rather2015recurrent,Sezeretal19,Ozbayoglu20,WeiC20,ItoSITY20,BiesnerRSLHLPLB22}. They employ multiple recurrent hidden layers to take the inputs from both the current states and the outputs of previous states' hidden layers to memorize historical long-range information and dependencies. Classic RNNs suffer from the vanishing gradient or exploding issues accomplished with their structural constraints. LSTM variants \cite{hochreiter1997long} further overcome the vanishing gradient issue by replacing neurons with novel memory cell structures. Specifically, three adaptive and multiplicative gating units (the input gate, the forget gate, and the output gate) compose such memory structures. As any errors backpropagated to the self-recurring unit are considerably uncontaminated, the vanishing gradient issue is abolished. LSTMs are commonly applied to model financial time series for market movement forecasting \cite{kim2018forecasting,namin2018forecasting}. In addition, recent progress includes deep state space models, deep factor model, and deep AR model \cite{rangapuram2018deep,salinas2020deepar}. These models do not involve cross-multivariate relation learning and do not model dependence structures. Our latest deep cross-multivariate net MTSNet \cite{Yang23} models cross-multivariates without involving dependence structures as well.

\subsection{Variational Deep Learning in Finance}

However, RNNs are built on point estimation with vulnerability and are incapable of capturing non-normal characteristics in financial markets including nonstationarity, non-IID features, distributional dynamics and overfitting to small-scale data \cite{Cao20-techniques,Serre19-dl}. A recent approach addresses some of these issues by combining RNNs with VAE and incorporating the VAE-based probabilistic estimation into RNN-based sequential modeling. For example, VRNN \cite{chung2015recurrent} combines VAE and RNN to capture the external nonlinear variations of financial volatilities. However, since a Gaussian process on latent variables is assumed to generate the corresponding conditional distribution of variables, it consequently fails to capture the asymmetric and nonlinear characteristics hidden in financial markets and suffers from the posterior collapse inherent in VAE. Other approaches \cite{miao2016neural} introduce new network architectures or changes to objective functions, with gaps remaining in jointly modeling complex market conditions and cross-market couplings. 

Another issue is the constrained mean-field assumption in VAE when copula functions capture high-dimensional dependencies in financial variables. Copula models for conducting the variational posterior to match the true posterior are introduced to generate copula-VAE structures for complex dependencies. For example, a copula variational inference uses a vine copula to estimate the empirical density of financial data \cite{tran2015copula}, and a multivariate Gaussian copula captures the joint distribution between variables \cite{wang2019neural}. The latter approach involves the reparametrization trick \cite{kingma2013auto} to estimate the joint distribution, which cannot capture the complex dependence structure with asymmetric and nonlinear characteristics using their multivariate Gaussian copula-based architecture. 

Our work further models complex cross-market conditions and couplings by integrating vine copula with variational neural learning.

\section{Preliminaries}
We introduce the background knowledge of copula methods, RNNs and VAE as the foundation of our method.

\subsection{Copula Methods}
\label{subsec:copula}

The copula methods \cite{Sklar1959,min2010bayesian,hafner2012dynamic,Elidan2013} are powerful in modeling  multivariate dependencies. Regardless as to whether they are  linear or nonlinear, complicated correlations between variables can be captured by appropriate copula functions. 
Following  Sklar's theorem \cite{Sklar1959}, the multivariate joint distribution $F$ of variables $x_1, \ldots, x_n$ with marginal distributions $F_1(x_1), \ldots, F_n(x_n)$ can be expressed as a copula function $C()$ with its marginal distributions $\{F_i\}$ ($F_i \in U(0,1)$) as variables. 
\begin{equation}
\label{copuladef}
F_1(x_1, x_2, \ldots, x_n)=C(F_1(x_1), \ldots, F_n(x_n))
\end{equation}
%$F$ is the joint distribution of a random vector $\mathbf{x}=[x_1, \ldots, x_n]$, and $C()$ is a $n$-dimensional copula function with uniformly distributed marginals $F_i \in U(0,1)$ on range $[0,1]$. 
%For any $n$-dimensional random vector $\mathbf{x}=(x_1, \ldots, x_n)$, given uniformly distributed marginals $F_i \in U(0,1)$ on range $[0,1]$ and the related copula function $C$, the corresponding joint distribution $F(x_1, \ldots, x_n)$ can be generated by Eqn. (\ref{copuladef}). 
Assume there is the inverse distribution function $F^{-1}_i$ of each marginal $\{F_i\}$, then the copula function in Eqn. (\ref{copuladef}) can be presented as:
%\begin{small}
\begin{equation}
\label{copulaCDF}
C(u_1, u_2, \ldots, u_n )=F(F_1^{-1}(u_1), \ldots, F_n^{-1}(u_n))
\end{equation}
%\end{small}
$u=F_n(x_n)$. Additionally, for a random variable $x_i$, it is always easier to obtain its density function than distribution function. Then, the density functions of copula $c$ are constitutionally introduced to join the density functions $f$ instead. Thus, for any continuous multivariate distribution $F$, with continuous and monotonically increasing densities $f_1(x_1), \ldots, f_n(x_n)$, we can infer the following from Eqn. (\ref{copuladef}):
%\begin{footnotesize}
\begin{equation}
\label{copuladen}
f(x_1, \ldots, x_n)=c(F_1(x_1), \ldots, F_n(x_n))\prod_{i=1}^nf_i(x_i)
\end{equation}
%\end{footnotesize}
$c$ is the density function of copula $C$. Per Eqn. (\ref{copulaCDF}), the joint density function can be separated by the copula density function(s), which describe the dependencies between marginal distributions and the density of marginal distributions. In terms of the above deduction, the selection of a copula does not rely on the selection of marginal distributions, and  marginal distributions could be diversified as well. Thus, the joint distribution of multivariates can be obtained by separately estimating their copula functions and marginal distributions.

Further, the vine theory \cite{bedford2002} presents copula dependence modeling based on graphical models. Let $V$, $T$, $E$ and $N$ represent the structure, trees, edges and nodes of a copula vine respectively, a regular vine and its associated concepts: \textit{complete union}, \textit{conditioning set}, \textit{conditioned set}, and \textit{constraint set}, given in \cite{bedford2002,Dissmann2013}, are essential for modeling high-dimensional dependence structures.

\begin{definition} \textbf{(Regular Vine)}
\label{Def_Regular_Vine}
$V = (T_1,...,T_{n-1})$ is a regular vine on $n$ elements if
\begin{enumerate}
    \renewcommand{\labelenumi}{(\theenumi)}
    \item $T_1$ is a tree with nodes $N_1 = {1,...,n}$ and a set of edges denoted by $E_1$;
    \item For $j =2,..., n-1$, $T_j$ is a tree with nodes $N_j = E_{j-1}$ and the edge set $E_j$;
    \item Proximity condition: for $j=2,...,n-1$ and ${a,b} \in E_j$,  $ \# (a \bigtriangleup b) =2$, where $\bigtriangleup$ denotes the symmetric difference operator and $\#$ denotes the cardinality of a set.	
\end{enumerate}
\end{definition}

\begin{definition}
\label{Def_Complete_Union}
\textbf{(Complete Union)} The complete union of an edge $e_j\in E_j$  is the set $U_{e_j}=\{n_1 \in N_1 \mid \exists e_k \in E_k, k=1,2,\ldots,j-1$ with $n_1 \in e_1 \in e_2 \in \ldots \in e_{j-1} \in e_j \} \subset N_1$.
\end{definition}

\begin{definition}
\label{Def_Cond_Set}
\textbf{(Conditioning and Conditioned Sets)} 
For $e_j=\{a,b\} \in E_j,a,b \in N_j, j=1,2, \ldots, n-1$, the conditioning set of an edge $e_j$ is $D_{e_j} = U_a \cap U_b$, and the conditioned sets of an edge $e_i$ are $C_{e_j,a} = U_a \setminus D_{e_j}, C_{e_j,b} = U_b \setminus D_{e_j}$ and $C_{e_j}=C_{e_j,a} \cup C_{e_j,b}=U_a \triangle U_b$, where $A \triangle B := (A \setminus B) \cup (B \setminus A)$ denotes the symmetric difference between two sets.
\end{definition}

\begin{definition}
\label{Def_Constraint_Set}
\textbf{(Constraint Set)} The constraint set for $V$ is a set:\\
\begin{equation}
%\centerline{$CV = \{(\{C_{e_a},C_{e_b}\},D_e) \mid e \in E_i, e = \{a,b\},i=1,\ldots,n-1\}$}
CV = \{(\{C_{e_a},C_{e_b}\},D_e) \mid e \in E_j, e = \{a,b\},j=1,\ldots,n-1\}
\end{equation}
\end{definition}

To avoid the restrictions on the vine specification and estimation, Bedford and Cooke \cite{bedford2002} proposed an approach based on the \textit{partial correlation} between variables. 
\begin{definition}\textbf{(Partial Correlation)}
\label{Dpar}
The partial correlation between two variables $x_1$ and $x_2$ given $x_3, \ldots, x_n$ is defined as:
\begin{equation}
%\small
\label{Epar}
\rho_{1,2;3,\ldots,n}=\frac{\rho_{1,2;3,\ldots,n-1}-\rho_{1,n;3,\ldots,n-1}\cdot\rho_{2,n;3,\ldots,n-1}}{\sqrt{1-\rho_{1,n;3,\ldots,n-1}^2}\cdot\sqrt{1-\rho_{2,n;3,\ldots,n-1}^2}}
\end{equation}
\end{definition}

The partial correlation ($\rho_{1,2}$) equals  Kendall's tau ($\tau_{1,2}$) between variables when removing the effect of the conditional variables. Thus, based on the above definition,  partial correlation can be easily achieved by Eqn. (\ref{Epar}) in an iterative process. Bedford and Cooke \cite{bedford2002} approved that, for any conditional distribution that follows an elliptical distribution, it equals the partial correlation between the variables. With this property, it is possible to estimate the conditional distribution on each node by calculating the equivalent partial correlation instead. 

In Section \ref{subsec:wprv}, the above concepts will be applied to construct our regular vine. A weighted partial regular vine captures the multivariate dependence degree and structure of cross-market financial variables. 

\subsection{RNN - LSTM}

RNNs apply deep sequential modeling structures suitable for time series data. At each time step of a time series, the inputs are transformed to hidden states which are further transformed as inputs to the next time step's hidden states. This forms the so-called recurrent architecture, enabling RNNs to sequentially model long-range dependencies and capture the dynamics of input time series. This makes RNNs more suitable for sequential modeling than CNNs.
%automatically, thus they are computationally more powerful than feed-forward networks, and the valuable approximation results are obtained for chaotic time series prediction. Although RNNs can provide well performance for the time series modeling, it is hard to learn long-term dependencies because of the vanishing gradient issue, as memory of older data may become subtle, causing the gradient signal to become very small.

A typical RNN model is  LSTM \cite{hochreiter1997long}, which sequentially models sequences with a long postponement and combination of components at different frequencies. To resolve the vanishing gradient issue  caused by the relatively ``long'' time using the recurrent architecture, LSTM introduces a novel memory cell structure consisting of three gates: input gate $i_t$, forget gate $f_t$, and output gate $o_t$, and a self-recurrent neuron. Interactions both within a memory cell and between it and its neighbouring memory cells are controlled by the gates. The output gate controls the state of the memory cell as to whether it can recast the states of other memory cells, and the forget gate decides to ``remember'' or ``forget'' the previous states. Each gate has the current input $\mathbf{x}_t$ and the previous hidden state $\mathbf{h}_{t-1}$ as inputs and has its own parameters ($\mathbf{U}$, $\mathbf{W}$ and $\mathbf{b}$), respectively. A general LSTM is formulated as follows \cite{hochreiter1997long}:
%\begin{small}
\begin{equation}
\label{EQLSTMDef}
\begin{split}
& \mathbf{f}_t =  \sigma(\mathbf{W}_f \cdot \mathbf{x}_t + \mathbf{U}_f \cdot \mathbf{h}_{t-1} + \mathbf{b}_f)  \\
& \mathbf{i}_t =  \sigma(\mathbf{W}_i \cdot \mathbf{x}_t + \mathbf{U}_i \cdot \mathbf{h}_{t-1} + \mathbf{b}_i)  \\
& \mathbf{o}_t =  \sigma(\mathbf{W}_o \cdot \mathbf{x}_t + \mathbf{U}_o \cdot \mathbf{h}_{t-1} + \mathbf{b}_o)  \\
& \mathbf{h}_t = \mathbf{o}_t \otimes \tanh{\mathbf{c}_t} \\
& \mathbf{\hat{c}}_t = \tanh{(\mathbf{W}_c \cdot \mathbf{x}_t + \mathbf{U}_c \cdot \mathbf{h}_{t - 1} + \mathbf{b}_c)} \\
& \mathbf{c}_t = (\mathbf{i}_t \otimes \mathbf{\hat{c}}_t) \oplus (\mathbf{f}_t \otimes \mathbf{c}_{t - 1}) \\
\end{split}
\end{equation}
%\end{small}
$\mathbf{c}_t$ and $\mathbf{h}_t$ are the final cell state and the hidden state for time $t$, while $\mathbf{\hat{c}}_t$ is the state vector of a new memory cell.

\subsection{Variational Autoencoder - VAE}
VAE is  composed of a probabilistic encoder $q_\phi$ with parameter $\phi$ and a probabilistic decoder $p_\theta$ with parameter $\theta$, and the encoding-decoding architecture similar to a classic autoencoder. It assumes that the marginal likelihood of input data can be estimated by a generative process on a set of samples i.i.d. drawn from a parameterized distribution with ground-truth generative factors $\mathbf{z}$ \cite{kingma2013auto}:
\begin{equation}
\label{EqVAE}
\mathbb{E}_{q_\phi(\mathbf{z}|\mathbf{x})} [\log{p_\theta}(\mathbf{x}|\mathbf{z})],
\end{equation}
%$\phi$ is the parameter of the distribution of the VAE's encoder $p_\theta$, $\theta$ parameterizes the distribution of the associated decoder $q_\phi$. 
Then, its related log-likelihood can be approximated by:
\begin{equation}
\label{EqVAELoss}
\log{p_\theta(\mathbf{x}|\mathbf{z})} = D_{KL}(q_\phi(\mathbf{z}|\mathbf{x})\|p_\theta(\mathbf{z})) + \mathcal{L}(\theta, \phi;\mathbf{x},\mathbf{z})
\end{equation}
where (1) $D_{KL}(\|)$ representing the non-negative Kullback-Leibler (KL) divergence between the true posterior $p_\theta(\mathbf{x}|\mathbf{z})$ and the approximate posterior $q_\phi(\mathbf{z}|\mathbf{x})$; and (2) the  variational lower bound $\mathcal{L}(\theta, \phi;\mathbf{x},\mathbf{z})$ applied on the marginal likelihood. 
%The objective of the learning process is to learn a posterior by minimizing the related KL divergence. 
With Eqn. (\ref{EqVAELoss}), minimizing the KL divergence can be alternatively obtained by maximizing the evidence lower bound (ELBO) $\mathcal{L}(\theta, \phi;\mathbf{x},\mathbf{z})$:
\begin{equation}
\label{EqELBO}
\begin{split}
\mathcal{L}(\theta, \phi;\mathbf{x},\mathbf{z}) =  & \mathbb{E}_{q_\phi(\mathbf{z}|\mathbf{x})}[\log{p_\theta(\mathbf{x}|\mathbf{z})}] - D_{KL}(q_\phi(\mathbf{z}|\mathbf{x})\|p(\mathbf{z})) \\
= & \mathbb{E}_{q_\phi(\mathbf{z})}[\log{p_\theta(\mathbf{x},\mathbf{z})}] - \mathbb{E}_{q_\phi(\mathbf{z})}[\log{q_\phi(\mathbf{z})}] \\
%- \mathbb{E}_{q_\phi(z)}[\log{q(z)}]
\end{split}
\end{equation}

Typically, mean-field approximation over samples ($i \in D$ with data $D$) simplifies the variational inference:
\begin{equation}
\label{EqMeanField}
q_\phi(\mathbf{z}) = \prod_{i}q_{\phi_i}(z_i)
%\prod^{D}_{i=1}q_\phi_i(z_i)
\end{equation}
which assumes the members of variational family $Q$ are dimensional-wise independent. However, a particular training difficulty called \textit{posterior collapse} arises along with the predigested variational family. Under this circumstance, the posterior yields to the prior, and the generative factor $p(\mathbf{x}|\mathbf{z})$ will no longer depend on the latent factors $\mathbf{z}$.

\section{The WPVC-VLSTM Model}

\begin{figure*}[t]
\centering
\includegraphics[scale=0.55]{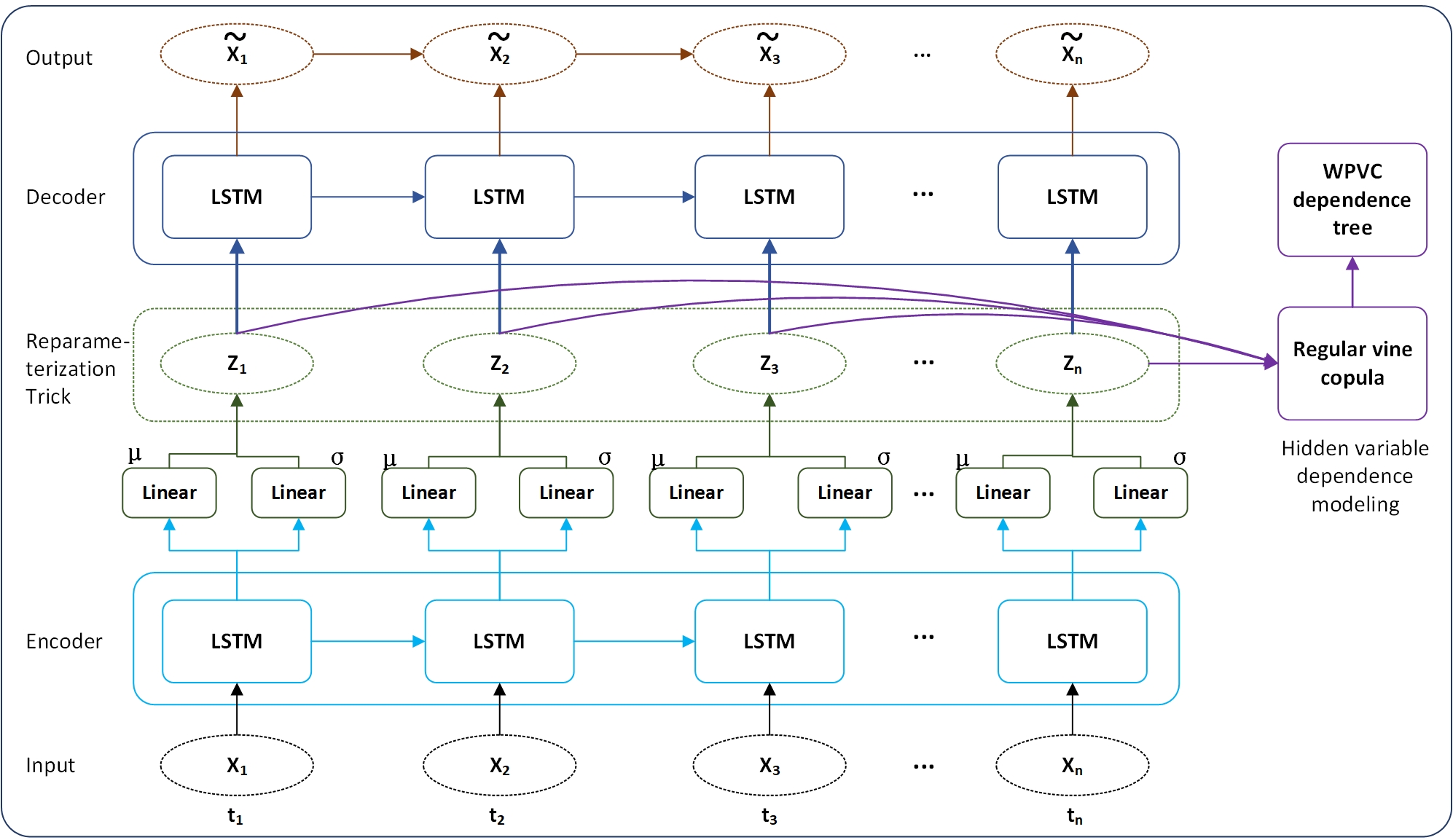}
\caption{WPVC-VLSTM: A weighted partial regular vine copula integrated into variational LSTM for modeling both cross-market dependence degrees and structures}
\label{FLSTM_VAE}
\end{figure*}

Figure \ref{FLSTM_VAE} shows the structure of the proposed variational LSTM (VLSTM) integrated with the weighted partial regular vine copula (WPVC) to model observable and hidden cross-market factors and their relations and structures. Below, we first hybridize VAE and LSTM to form VLSTM characterizing the temporal hidden dependencies between financial time series variables. WPVC then captures the dependencies between the latent variables in VAE to characterize the dependence degrees and structures between financial variables. Lastly, we explain the WPVC construction and its vine optimization based on partial correlation. 

\subsection{LSTM-based Variational Autoencoder}
\label{subsec:vlstm}

General RNNs are inadequate to capture the convoluted volatility between finance markets. Inspired by \cite{chung2015recurrent}, we integrate  VAE into LSTM by introducing latent variables $\mathbf{z}$ to parameterize the joint distribution of inputs $\mathbf{x}$. With the reparameterization trick  \cite{kingma2013auto}, hidden states are assumed to follow a normal distribution with mean $\mu$ and variance $\sigma$. Thus, the posterior can be represented as a deterministic function with parameters $\mu$ and $\sigma$ and the latent code can be consequently generated from the posterior by sampling. 

In Fig. \ref{FLSTM_VAE}, given input $\mathbf{x}_{t}$ at time $t$, the encoder approximates posterior $p_{\theta}(\mathbf{z}_{t} | \mathbf{x}_{t})$ by feeding LSTM outputs into two linear modules to estimate parameters $\mu$ and $\sigma$ of the latent variables. Further, the randomly sampled $\mathbf{z}$ from  posterior $p_{\theta}(\mathbf{z}_{t} | \mathbf{x}_{t})$ is fed into the decoder's LSTM. The generative model is described as:
\begin{equation}
\label{LSTM_VAE1}
\begin{split}
p_{\theta}(\mathbf{x}_{t} | \mathbf{z}_{\leq t}, \mathbf{x}_{\leq t}) = \mathcal{L}(h^{\mu, \sigma}(g_z(\mathbf{z}_t), \mathbf{h}_{t - 1})) \\
q_{\phi}(\mathbf{z}_t | \mathbf{z}_{\leq t}, \mathbf{x}_{\leq t}) = \mathcal{L}(h^{\mu, \sigma}(g_x(\mathbf{x}_t), \mathbf{h}_{t - 1})) \\
\end{split}
\end{equation}
The volatility item $\sigma_{x,t}$ is epitomized in the hidden state $\mathbf{h}$, $g_x$ and $g_z$ are neural networks to extract features from $\mathbf{x}_t$ and $\mathbf{z}_t$ respectively. The parameterization of  VLSTM is as follows:
\begin{equation}
\label{LSTM_VAE2}
p_{\theta}(\mathbf{x}_{\leq T}, \mathbf{z}_{\leq T}) = \prod^{T}_{t = 1}p_{\theta}(x_t | \mathbf{z}_{\leq t}, \mathbf{x}_{\leq t})p_{\theta}(z_t | \mathbf{z}_{< t}, \mathbf{x}_{< t})
\end{equation}
$p(\mathbf{x}_{\leq T}, \mathbf{z}_{\leq T})$ is the joint distribution between all observations up to time $T$. Then, the hidden states of VLSTM are updated after the generation of $\mathbf{z}_t$ and $\mathbf{x}_t$ at every time $t$.
\begin{equation}
\label{LSTM_VAE3}
\mathbf{h}_t = f_\tau(g_x(\mathbf{x}_t), g_z(\mathbf{z}_t), \mathbf{h}_{t - 1})
\end{equation}
$f$ denotes the forget gate in the LSTM cell as described in Eqn. (\ref{EQLSTMDef}), and $\tau$ is the parameter set of $f$. The posterior is a function determined by both $\mathbf{x}_t$ and $\mathbf{h}_{t-1}$, which is intractable. Hence, the approximate posterior $q_\phi(\mathbf{z}_{\leq T} | \mathbf{x}_{\leq T})$ can be approached by the time step-wise EBLO:
%\begin{small}
\begin{equation}
\begin{split}
\label{EQ_VLSTM_ELBO}
\mathcal{L}_{VAE} =  & \mathbb{E}_{q_\phi(\mathbf{z} \leq T| \mathbf{x} \leq T)}[\sum_{t = 1}^{T}(\log{p_\theta(\mathbf{x}_t | \mathbf{z}_{\leq t}, \mathbf{x}_{< t})} \\
& - D_{KL}(q_\phi(\mathbf{z}_t | \mathbf{x}_{\leq t}, \mathbf{z}_{< t}) \| p_\theta(\mathbf{z}_t | \mathbf{x}_{< t}, \mathbf{z}_{< t})))] \\
\end{split}
\end{equation}
%\end{small}

For each LSTM cell, a fully connected layer is introduced to generate the model's output $\mathbf{y}_t = g_y(\mathbf{h}_t)$ at each time stamp $t$. For an accurate forecasting over the existing time series, the loss between the true output $\mathbf{y}$ and the corresponding predicted $\mathbf{\hat{y}}$ should be minimized. Therefore, the cross-entropy between $\mathbf{y}_t$ and $\mathbf{\hat{y}}_{t}$ is introduced to express the loss of prediction:
\begin{equation}
\label{EQ_VLSTM_OUT}
\mathcal{L}_{P} = \sum_{t=1}^{T}-p_\theta(\mathbf{y}_t)\log{q_\phi(\mathbf{\hat{y}}_t)}
\end{equation}

With the structure of our model, the ELBO of VAE $\mathcal{L}_{VAE}$ and the prediction loss $\mathcal{L}_{P}$ can be employed to constrain the latent code $\mathbf{z}$ and reduce the influence caused by inessential variables during optimization. We thus obtain the comprehensive loss function $\mathcal{L}_{VLSTM}$:
\begin{equation}
\label{EQ_LOSS_VLSTM}
\mathcal{L}_{VLSTM} = \mathcal{L}_{P} - \mathcal{L}_{VAE}
\end{equation}

To estimate the parameters, VLSTM firstly moves along the track to maximize expectation $\mathbb{E}_{q_\phi(\mathbf{z}|\mathbf{x})}[p_\theta(\mathbf{x}|\mathbf{z})]$. There would be a congenital ``frontier'' for  posterior $q_\phi(\mathbf{z}|\mathbf{x})$ where the expectation cannot increase anymore, since the posterior will never attain the true posterior $p_\theta(\mathbf{x}|\mathbf{z})$. Then, the process turns to seek the minimized  KL divergence to achieve the maximized ELBO. The posterior will finally yield to the prior with no adequate gradient to move due to the maximized ELBO. Then the process cannot take the advantages from minimizing the KL divergence any more. In this case, if the mean-field assumption with a factorized form can be substituted by introducing additional dependence factors (more specifically, a copula) between variables for the variational family, due to the elastic form of the distribution of posterior $q$, the true posterior $p$ will be more attainable by maximizing the likelihood. Therefore, we construct a weighted partial regular vine copula (WPVC) to address this issue below.

\subsection{Weighted Partial Regular Vine Construction}
\label{subsec:wprv}

Minimizing the loss function $\mathcal{L}_{VLSTM}$ in Eqn. (\ref{EQ_LOSS_VLSTM}) desires a proper factorization of the variational family of distributions instead of the mean field independent assumption. As discussed in Section \ref{subsec:copula}, the vine copula takes advantage of bivariate copulas capable of expressing a broad range of dependencies satisfying this requirement. We consequently employ the vine copula to describe the joint distribution of posterior in VLSTM in Section \ref{subsec:vlstm}. For a given posterior, a copula to approximate the joint posterior can be proposed as follows:
\begin{equation}
\label{EQ_Copula_Post}
q_\phi(\mathbf{z} | \mathbf{x}) = c(\mathbf{u}; \mathbf{\eta})\prod_{t = 1} ^ {T}q_\phi(\mathbf{z}_t | \mathbf{x}_{\leq t}, \mathbf{z}_{< t}; \mathbf{\lambda})
\end{equation}
where $\mathbf{u} = (u_1,...,u_t)^T$, $u_t$ refers to the marginal distribution with density $q_{\phi_t}$, $\mathbf{\lambda}$ and $\mathbf{\eta}$ refer to the mean-field and copula parameters, respectively.

To avoid specifying the marginal distribution of $\mathbf{z}$, a nonparametric vine without structure restrictions needs to be constructed. An appropriate vine structure should keep the weakest correlation on the bottom tree and ensure the tree on the top owns own the strongest correlation. However, most of the top-to-bottom vine construction approaches imply strong assumptions on vine structures. As per the approach in \cite{bedford2002} for the vine structure construction based on partial correlation, a partial correlation for elliptical distributions is equivalent to its corresponding conditional correlation. Accordingly, a nonparametric bottom-to-top strategy can be employed for regular vine construction by iteratively estimating the partial correlation on each node. This approach ensures that the weakest and strongest correlations can stick to the bottom and the top trees respectively; and the trees on other levels do not depend on the structure of its previous trees. By leveraging the regular vine, the resultant factorization is spontaneously more malleable to capture the high-dimensional dependencies of latent codes.

%\begin{enumerate}
%\renewcommand{\labelenumi}{(\theenumi)}
%  \item There are $(j -1)$ and $(j +1)$ ($j+1 \le n$) variables in the conditioning sets and constraint sets of an edge of the $j^{th}$ tree respectively;
%  \item If two or more nodes have the same constraint sets, they are the same node;
%  \item If variable $z_i$ is a member of the conditioned set of an edge $e$ in a regular vine, then $z_i$ is a member of the conditioned set of exactly one of the m-children of $e$, and the conditioning set of the m-child is a subset of $De$.
%  \item If two or more nodes have the same constraint sets, they are the same node;
%  \item If variable $z_i$ is a member of the conditioned set of an edge $e$ in a regular vine, then $z_i$ is a member of the conditioned set of exactly one of the m-children of $e$, and the conditioning set of the m-child is a subset of $D_e$.
%\end{enumerate}

\begin{figure}[t]
\centering
\includegraphics[scale=0.3]{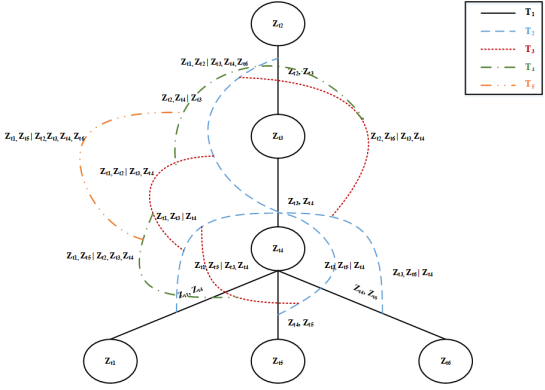}
\caption{Regular vine tree modeling the dependencies between LSTM temporal variables: illustration of six variables $\mathbf{z}$ for time points $t_1$ to $t_6$.}
\label{FRVine}
\end{figure}

VLSTM constructs an appropriate regular vine structure to capture the comprehensive dependencies between variables in latent codes $Z$. To this end, we introduce two lemmas similar to that in \cite{xu2017copula} on top of some important properties in \cite{kurowicka2006} to identify the nodes and their m-children for each tree by an iterative process in the regular vine.
\begin{lemma}
\label{L1}
Let $I \in \{1,...,n\}$, $z_{t_1},z_{t_2}, z_{t_3},z_{t_4} \in I$ and $z_{t_1} \neq z_{t_2}$, the nodes of the tree $T_j$  be $N_1 = \{z_{t_1},z_{t_3} \, ; \,I \backslash \{z_{t_1},z_{t_2}, z_{t_3}\}\}$ and $N_2 = \{z_{t_2},z_{t_4} \, ;  \, I  \backslash \{z_{t_1},z_{t_2},z_{t_4}\}\}$. Nodes $N_1$ and $N_2$ share a common m-child within the regular vine. If $z_{t_3} \neq z_{t_4}$, the common m-child is $\{z_{t_3},z_{t_4} \, ; \, I \backslash \{z_{t_1},z_{t_2},z_{t_3},z_{t_4} \} \}$.
\end{lemma}

\begin{lemma}
\label{L2}
Let $V$ be a regular vine composed of $n$ random variables $\{z_n\}$. For trees $T_2,...,T_{n-1}$ in $V$, every edge in tree $T_j$ (where $j = 2,...,n-1$) has and only has two constraint sets of m-children in tree $T_{j-1}$, and these two constraint sets are generated from the same conditioned set with different indexing variables.
\end{lemma}

The objective of constructing the regular vine for VLSTM is to build a number of trees that keep the weakest correlation at the bottom and ensure the tree on top of the vine holds the strongest correlation. However, when identifying the appropriate candidate combination for a specific tree at each level, we can only choose either the strongest or weakest correlation.
Therefore, we also need to introduce an additional parameter $l$ (called the tree inverse indicator) to indicate the level of the vine for the selection of the weakest rather than the strongest correlation. More specifically, assume $j = 1,...,n-1$ denotes the level of the trees in a regular vine, for trees ($T_{n-1},...,T_{l-1}$) sit underneath the inverse level ($k < l$), the selected combination of candidate nodes seek to minimize the function $\sum|\rho_{c;d}|$, where $c$ and $d$ are the \textit{conditioning} and \textit{conditioned} sets as per Definition \ref{Def_Complete_Union}. For the trees ($T_{l},...,T_1$) sitting at or above the inverse level ($k \geqslant l$), the selected nodes need to maximize the function $\sum \log{(1 - \rho_{c;d}^2)}$ conversely. 

%To keep the weakest correlation on the bottom and the strongest correlation on the top, the key step is to identify the appropriate nodes for each candidate combination. We introduce a tree inverse level $k$ to identify the selection of the weakest or strongest correlation at each level. More specifically, assume $l$ is the level of the trees in a regular vine structure, for trees under the inverse level ($l \geqslant k$), the appropriate nodes must minimize the value of function $\sum|\rho_{c;d}|$, where $c$ and $d$ are the \textit{conditioning} and \textit{conditioned} sets defined in Definition \ref{Def_Complete_Union}. Additionally, for the trees beyond the inverse level ($l < k$), the appropriate nodes must maximize the value of function $\sum \log{(1 - \rho_{c;d}^2)}$.

Below, we further demonstrate the process of building an optimal vine structure on the latent variables of VLSTM. Given the latent variable set $Z_t$ at time $t$, assume the latent variables contain six features denoted by $z_{t_1}$, $z_{t_2}$, $z_{t_3}$, $z_{t_4}$, $z_{t_5}$ and $z_{t_6}$, and the tree inverse indicator $l$ is set to  three. As per the properties of a regular vine, a vine structure thus consists of $5$ trees and $20$ nodes, each node containing either a bivariate copula or a partial correlation. The overall vine structure is shown in Fig. \ref{FRVine}.

%Below, we illustrate the construction of an optimal vine structure for the hidden variables of the VLSTM. Given the hidden variable set $Z_t$ at time $t$, assume there are five features denoted by $z_{t_1}$, $z_{t_2}$, $z_{t_3}$, $z_{t_4}$ and $z_{t_5}$ associated with the set at different time steps, and the tree inverse level $l$ is equal to 3. The regular vine then consists of $4$ trees and $14$ nodes in the regular vine structure based on partial correlation. All the trees and nodes are shown in Figure \ref{FRVine}. Each node can be allocated to one bivariate copula or one partial correlation.

Assume the weakest partial correlation is $\rho_{z_{t_1},z_{t_5};z_{t_2},z_{t_3},z_{t_4}, z_{t_6}}$ among the total of $21$ partial correlations composed of these six variables. Accordingly, $\{z_{t_1},z_{t_5}\}$ and $\{z_{t_2},z_{t_3},z_{t_4}, z_{t_6}\}$ are the conditioned set and conditioning set, respectively. 
Per Lemma \ref{L2}, which avoids the circumstance that the structure cannot be specified and maintains the regularity of the regular vine tree, the two candidate variables $\{z_{t_1},z_{t_5}\}$ in the conditioned set are firstly assigned to two different constraint sets together with the conditioning set $\{z_{t_2},z_{t_3},z_{t_4}, z_{t_6}\}$. Then the two nodes on tree $T_5$ own the constraint sets $\{z_{t_1},z_{t_2},z_{t_3},z_{t_4},z_{t_6}\}$ and $\{z_{t_1},z_{t_2},z_{t_3},z_{t_4},z_{t_5}\}$, respectively. As the regular vine for VLSTM takes a bottom-to-top tree construction approach, the weakest correlation is associated with the tree at the bottom. 

Assume that from all possible partial correlations, the combinations $\rho_{z_{t_1},z_{t_2};z_{t_3},z_{t_4},z_{t_6}}$ and $\rho_{z_{t_1},z_{t_5};z_{t_2},z_{t_3},z_{t_4}}$ reach the smallest absolute value. Then, the nodes of tree $T_5$ can be specified as $\{\{z_{t_1},z_{t_2}\},\{z_{t_3},z_{t_4},z_{t_6}\}\}$ and $\{\{z_{t_1},z_{t_5}\},\{z_{t_2},z_{t_3},z_{t_4}\}\}$, and the node in the middle of $T_4$ can also be specified as $\{\{z_{t_1},z_{t_5}\},\{z_{t_3},z_{t_4}\}\}$ per Lemma \ref{L1}. The constraint sets on the nodes which sit on the left and right hand sides of $T_4$ are correspondingly $\{z_{t_1},z_{t_2},z_{t_3},z_{t_4}\}$ and $\{z_{t_2},z_{t_3},z_{t_4},z_{t_6}\}$. As we still need to identify the combination with the smallest correlation on this level, assume the combination with the weakest partial correlation among all candidates from the constraint sets are $\rho_{z_{t_1},z_{t_5};z_{t_2},z_{t_3},z_{t_4}}$,  $\rho_{z_{t_1},z_{t_2};z_{t_3},z_{t_4},z_{t_6}}$ and $\rho_{z_{t_2},z_{t_6};z_{t_3},z_{t_4}}$, we will obtain the two border nodes $\{\{z_{t_1},z_{t_2}\},\{z_{t_3},z_{t_4}\}\}$ and $\{\{z_{t_2},z_{t_6}\},\{z_{t_3},z_{t_4}\}\}$. Similar to the process described for the $T3$ construction, we will still achieve a line-like structure with  nodes $\{\{z_{t_1},z_{t_3}\},\{z_{t_4}\}\}$ and $\{\{z_{t_3},z_{t_5}\},\{z_{t_4}\}\}$ in the middle, and  $\{\{z_{t_2},z_{t_3}\},\{z_{t_4}\}\}$ and $\{\{z_{t_3},z_{t_4}\},\{z_{t_6}\}\}$ on both sides. A similar process can be employed for trees $T_2$ and $T1$ as well. In accordance with the specified constraint sets on this level, $T2$ has a chance to be split into a star-like structure. Additionally, since the tree inverse indicator equals three, the candidate combination with the strongest rather than the weakest partial correlation needs to be selected for border nodes on this level. Then  tree $T_1$ on the top level of the vine is specified along with the specification of $T_2$ due to structure restrictions. Consequently, the whole vine structure is specified.

With the above bottom-to-top regular vine construction process, as the tree inverse indicator $k$ can be selected from $n - 1$ to $\lceil \frac{n}{2} \rceil$, there are a total of $\lceil \frac{n}{2} \rceil$ possible regular vine structures that can be generated using different settings of the value for $k$. 

The next step is to determine the `best' vine structure from these candidates. As previously discussed, the `best' regular vine structure should keep weak correlations as low as possible, and float up  strong correlations as high as possible. To address this requirement, for two variables $z_i$ and $z_i'$, we define a deterministic function composed of the weighted determinant for our regular vine structure selection as follows:
\begin{equation}
\label{DefBestD}
R = -\log{\prod_{z_i,z_i'}(1 - W_i\rho_{z_i, z_i';r(z_i, z_i')} ^ 2)}
\end{equation}
$r(z_i,z_i')$ denotes the conditioning set excluding variables $z_i$ and $z_i'$, and $W_i$ is the corresponding weight matrix. The vine structure that can maximize Eqn. (\ref{DefBestD}) will be subsequently selected. The algorithm for the above partial regular vine construction is demonstrated in Algorithm \ref{Alg_Rvine_B2T}.

\begin{algorithm}
\caption{Weighted Partial Regular Vine Construction}
\begin{algorithmic}[1]
%\label{A1}
\REQUIRE A random vector with $n$ variables\\
%\ENSURE~~ \\
%Classifiers: $(f_1, f_2,\ldots, f_m)$.\\
\STATE Evaluate partial correlation for all the possible combinations from input variables, find out the pair with the smallest absolute value and assign it to the node in the bottom tree $T_{n-1}$.
\FOR {$l = 1, \ldots, n - 2$}
    \FOR {$j = n - 1, \ldots, \lceil \frac{n}{2} \rceil$}
        \IF {$T_j < T_l$}
            \STATE Calculate the partial correlation for all possible pairs, and minimize  function $|\rho_{c:d}|$ for tree $T_j$.
        \ELSE
            \STATE Calculate the partial correlation for all possible pairs, and maximize  function $\sum \log{(1 - \rho_{c:d} ^ 2)}$ for  tree $T_j$.
        \ENDIF
    \ENDFOR
\STATE Keep the vine structure and denote it as $V_l$.
\ENDFOR
\STATE A set of partial regular vines $V_1,\ldots, V_{n - 2}$ is specified.
\STATE Calculate  function $-\log{(R)}$ for all candidate partial regular vines, and select the vine structure to be denoted as $V$ that can maximize the function as the `best' vine structure.
\RETURN The selected partial regular vine structure $V$.
\end{algorithmic}
\label{Alg_Rvine_B2T}
\end{algorithm}

\subsection{Vine Structure Optimization and Copula Family Selection}

With the increase of input dimensionality, the number of parameters included in the joint distribution that need to be estimated increases exponentially. Given a vine composed of $n$ variables, if the \textit{t}-copula contains two parameters for each node, the total number of parameters included in the decomposition of the joint distributions will be $n(n-1)$. The computational cost of parameter estimation will be huge with the increase of input variables. Therefore, an appropriate truncation for the vine copula is crucial as well. As discussed in \cite{Aas2009}, $C_{e(a),e(b) \,| \, L_e}^{independence} = 1$ due to the conditional independent copula. Thus, the conditional bivariate independent copula can be employed to substitute those nodes with very weak correlations. This methodology ensures that the significant dependencies  derived by strong correlations are kept and the less important dependencies from weak correlations are removed. Specifically, a truncation threshold $\rho_{trun}$ which lies between zero and one can be introduced to indicate which partial correlations need to be truncated. If the partial correlation on one node that is less than the truncation threshold $\rho_{trun}$, it will be replaced by a conditional independent copula, then the whole regular vine is consequently abridged.

%\begin{algorithm}[h!] \small
%\caption{Partial Regular Vine Tree Structure Truncation}
%\begin{algorithmic}[1]
%\REQUIRE The $n$-variable partial canonical vine tree structure $V_{\rho_e}^R$ and truncation value $\rho_{trun}$
%\STATE Calculate the corresponding partial correlations based on the partial regular vine tree structure $V_{\rho_e}^R$;
%\FOR {$j=1,...,n-1$}
%\STATE In tree $T_j$, find all edges for which the absolute value of partial correlation is less than $\rho_{trun}$, e.g. $\rho_{e(p), e(q) \, ; \,  D(e)} < \rho_{trun}, e=\{p,q\}$ ;
%\STATE For those edges found in Step 4, replace all bivariate copulas with conditional independence copulas;
%\ENDFOR
%%\STATE For the 'best' canonical vine, the small absolute values of partial correlation, which are less than significance value $\rho$, are set to zero.
%%\STATE The optimal canonical vine based on conditional copula is corresponding to the canonical vine based on partial correlation.
%\RETURN The truncated partial regular vine tree structure
%\end{algorithmic}\label{AlgTrun}
%\end{algorithm}

%\label{Sec_CopSel}
Once the whole vine structure is specified, the next step is to assign an appropriate copula function for each node. It is remarkable that as the property of a partial correlation equaling its conditional correlation is limited to elliptical distributions, the bivariate copula on each node in the partial regular vine theoretically should be selected from elliptical copula families, such as the Gaussian or \textit{student-t} copula. However, Bedford proved in \cite{bedford2002} that there exists a unique mapping relationship between the regular vine constructed by partial correlation and correlation matrices. With this theorem, the selection of the bivariate copula on each node in the partial regular vine can be extended from the limited elliptical copula family to all possible bivariate copulas to empower the regular vine capable of describing the distributions with asymmetric and tail dependencies. 

%\newtheorem{theorem}{Theorem}
% \begin{theorem}
% \label{Theorem_PartialMap}
% For any regular vine on $n$ variables, there is a one-to-one correspondence between the set of $n \times n$ positive definite correlation matrices and the set of partial correlation specification of the vine.
% \end{theorem}

%The proof of Theorem \ref{Theorem_PartialMap} can be referred to \cite{bedford2002}, which is omitted here. The theorem shows that there is a one-to-one relationship between the partial regular vine specification and the correlation matrix, which ensures that we can map our  partial regular vine tree structure to the typical conditional correlation-based regular vine tree structure. We can then choose bivariate copulas from a large number of copula family candidates rather than just the elliptical copula family. Hence, the above limitation can be removed in selecting the bivariate copulas.

\section{Variational Inference and Parameter Estimation}

\subsection{Inference of Copula Variational LSTM}
With Eqn. (\ref{EQ_Copula_Post}), the log-likelihood for the joint posterior is:
%\begin{small}
\begin{equation}
\label{EQ_Copula_Likelihood}
\log{q_\phi(\mathbf{z} | \mathbf{x})} = \log{c(\mathbf{u}; \mathbf{\eta})} + \sum_{t = 1}^{T}\log{q_\phi(\mathbf{z}_t | \mathbf{x}_{\leq t}, \mathbf{z}_{< t}; \mathbf{\lambda})}
\end{equation}
%\end{small}
Inspired by the inference to independently estimate the marginal distribution and the copula function presented in \cite{joe2005asymptotic}, a two-step inference process is employed to estimate the parameters for the factors on the right-hand side of the above equation. Initially, the marginal distribution deals with the reparameterization trick \cite{kingma2013auto}. By this transformation, the marginal distribution is approximated as a deterministic function composed of two parameters: $\mu$ and $\sigma$, then the deterministic function is given by:
\begin{equation}
\label{EQ_Marginal_Distribution}
\mathbf{z} \sim \mu + \sigma\cdot\epsilon
\end{equation}
where $\epsilon \sim \mathcal{N}(0, I)$. Thus, we have $\prod_t q_\phi(z_t | x_{\leq t}, z_{<t}) \sim \mathcal{N}(\mu, \sigma^2I)$, and the sum of the logarithm of the marginal distributions (the mean field part) is given by:
\begin{small}
\begin{equation}
\label{EQ_MeanField_Log}
\begin{split}
& \sum_{t = 1} ^ {T}\log{q_\phi(z_t | x_{\leq t}, z_{<t})} \\
= & -\sum_{t = 1} ^ {T}\log{|\sigma_t|} - \sum_{t = 1} ^ {T}\frac{(z_t - \mu_t)^2}{2\sigma^2_t} - \frac{d}{2}\log{2\pi}
\end{split}
\end{equation}
\end{small}

Following the reparameterization trick, the gradients propagate inside the expectation. On the other hand, with our proposed vine copula variational family, the terms $\nabla_{\mathbf{z}_t}\log{q_\phi(\mathbf{z}; \mathbf{\mu}, \mathbf{\sigma}, \mathbf{\eta})}$ and $\nabla_{u_t}\mathbf{z}(u; \mathbf{\mu}, \mathbf{\sigma}, \mathbf{\eta})$ need to be calculated for each time stamp $t$. To simplify the formulas, the conditions of  latent variable $\mathbf{z}$ are denoted by $\mathbf{\psi}$, then the former can be decomposed as:
\begin{small}
\begin{equation}
\label{EQ_Gradient1}
\begin{split}
& \nabla_{\mathbf{z}_t} \log{q_\phi(\mathbf{z} | \mathbf{\psi}; \mathbf{\mu}_t, \mathbf{\sigma}_t, \mathbf{\eta}_t)} \\
= & \nabla_{\mathbf{z}_t} \log{q_\phi(\mathbf{z}_t | \mathbf{\psi}_t; \mathbf{\mu}_t, \mathbf{\sigma}_t)} \\
  & + \nabla_{ F(\mathbf{z}_t | \mathbf{\psi}_t; \mathbf{\mu}_t, \mathbf{\sigma}_t)} \log{c(F(\mathbf{z} | \mathbf{\psi}; \mathbf{\mu}, \mathbf{\sigma}); \eta)} \nabla_{\mathbf{z}_t} F(\mathbf{z}_t | \mathbf{\psi}_t; \mathbf{\mu}_t, \mathbf{\sigma}_t) \\
= & \nabla_{\mathbf{z}_t} \log{ q_\phi(\mathbf{z}_t | \mathbf{\psi}_t; \mathbf{\mu}_t, \mathbf{\sigma}_t)} \\
  & + q_\phi(\mathbf{z}_t | \mathbf{\psi}_t; \mathbf{\mu}_t, \mathbf{\sigma}_t) \sum_{j=1}^{d-1} \sum\nabla_{ F(\mathbf{z}_t | \mathbf{\psi}_t; \mathbf{\mu}_t, \mathbf{\sigma}_t)} \log{c_{kl|D_{e}}} \\
\end{split}
\end{equation}
\end{small}

The summation on the right-hand side of the above equation contains all bivariate copulas decomposed by our partial regular vine with  argument $F(\mathbf{z}_t | \mathbf{\psi}_t; \mathbf{\mu}_t, \mathbf{\sigma}_t)$. The next step is to infer  parameters $\eta$ for the copula part, which is a partial regular vine. Assume all the bivariate copulas within the vine are differentiable with respect to the corresponding parameters,  the reparameterized gradient for the VAE in the loss function can be written as:

\begin{equation}\label{EQ_Gradient2}
\begin{split}
\nabla_{\eta} \mathcal{L}_{VAE} = & \mathbb{E}_{\Sigma(u)} [\nabla_{\mathbf{z}} \mathbb{E}_{q_\phi(\mathbf{z} | \mathbf{x})} \sum_{t = 1}^{T} \log{p_\theta(\mathbf{x}_t | \mathbf{\psi}_t)} \\
& - \nabla_{\mathbf{z}} \mathbb{E}_{q_\phi(\mathbf{z} | \mathbf{x})} (\log q_\phi(\mathbf{x},\mathbf{z}) \\ 
& - \log{p_\theta(\mathbf{z})}) 
 \nabla_{u_t} \mathbf{z} (u; \mathbf{\mu}, \mathbf{\sigma}, \mathbf{\eta}) \\
& + \nabla_{\mathbf{z}} \mathbb{E}_{q_\phi(\mathbf{z} | \mathbf{x})} \log{c(u_1, \ldots, u_T)} \nabla_{u_t} \mathbf{z}(u; \mathbf{\mu}, \mathbf{\sigma}, \mathbf{\eta})] 
\end{split}
\end{equation}

As  entropy $\mathcal{L}_{P}$ does not depend on the parameters in $\mathcal{L}_{VAE}$, the parameters in Eqn. (\ref{EQ_VLSTM_OUT}) can be estimated by minimizing the entropy at last once the variational part is optimized. The whole process can be implemented by a generalized method of moments applying the expectation maximization (EM) algorithm within an iterative procedure. In each step, the objective function of variational inference applying the partial regular vine increases monotonically and consequently achieves a better approximation for the posterior distribution. A summary of the entire WPVC-VLSTM inference process is outlined in Algorithm \ref{ALG_VLSTM_Inference}.

\begin{algorithm}[t]
\caption{Variational LSTM Inference for WPVC-VLSTM}
\begin{algorithmic}[1]
\REQUIRE Time series variables $\mathbf{x}$ and the  output $\mathbf{y}$
\STATE Set up the iteration count, batch size for parameter estimation, and a threshold $\delta$ for loss function $\mathcal{L}_{VLSTM}$.
\STATE Initialize the parameters $\lambda$, $\eta$.
\WHILE{$\mathcal{L}_{VLSTM} \geqslant \delta$} 
%\COMMENT{for each iterator}
  \FOR {each batch}
    \STATE Generate the latent variable $\mathbf{z}$ as per Eqn. (\ref{LSTM_VAE1}) using the input $\mathbf{x}$.
    \STATE Generate the reconstructed $\mathbf{\hat{x}}$ by importing  $\mathbf{z}$ into the VLSTM decoder.
    \STATE Update $\mathbf{\lambda} = \mathbf{\lambda} + \rho_t\nabla_\mathbf{\lambda}\mathcal{L}_{VAE}$ by Eqn. (\ref{EQ_Gradient1}) until convergence
    \STATE Update $\mathbf{\eta} = \mathbf{\eta} + \rho_t\nabla_\mathbf{\lambda}\mathcal{L}_{VAE}$ by Eqn. (\ref{EQ_Gradient2}) until converged
    \STATE Generate  output $\mathbf{\hat{y}}$ and estimate the parameters by minimizing  Eqn. (\ref{EQ_VLSTM_OUT})
  \ENDFOR
%  \IF {$\mathcal{L}_{VLSTM} < \delta$}
%    \STATE break
%  \ENDIF
%\ENDFOR
\ENDWHILE
\RETURN A trained weighted partial regular vine variational LSTM Model WPVC-VLSTM
\end{algorithmic}
\label{ALG_VLSTM_Inference}
\end{algorithm}

\subsection{Computational Complexity Analysis}
For a partial regular vine with $n$ latent variables, there are a total of $n(n - 1)/2$ partial correlations which need to be evaluated in the vine construction process. The corresponding full decomposition involves the complexity of $\mathcal{O}(n^2)$. As we introduce a truncated low-rank copula decomposition by replacing weaker correlations, if the number of bivariate copulas reduces to $hn$ for $h>0$, the computational complexity falls to $\mathcal{O}(hn)$. In the decomposition of our copula function, there are $hn$ bivariate copulas derived from the final truncated partial regular vine. Consequently, the complexity of the calculation on the stochastic gradients of the parameters for the marginal distributions ($\mathbf{\lambda}$) and the copulas ($\mathbf{\eta}$) is $\mathcal{O}(hn)$ as well.

\section{Experiments}
Here, we introduce the data sets, baseline methods, model specifications, and evaluation metrics for the evaluation.

\subsection{Date Sets}
To investigate the effectiveness of WPVC-VLSTM, real data across different financial markets include 16 financial variables. The portfolio consists of (1) eight important comprehensive indices in the world: \^{}GSPC, \^{}STOXX, \^{}FTSE, \^{}FCHI, \^{}BFX, \^{}N225, \^{}HSI, and \^{}AEX; and (2) eight important currencies against the US dollar: EUR/USD, GBP/USD, CHF/USD, JPY/USD, HKD/USD, SGD/USD, CAD/USD, and AUD/USD. All  data were extracted from Yahoo Finance\footnote{http://finance.yahoo.com/}. The training dataset consists of 780 daily returns observed from 04 Jan 2016 to 28 Dec 2018. 246 daily returns from 07 Jan 2019 to 27 Dec 2019  are used for out-of-sample validation. 

\subsection{Baseline Methods}

A number of baselines implemented by different modeling techniques are used for comparison, including a classic statistical model, a stochastic volatility model, and three DNN models CNN, LSTM and variational LSTM models. We customize these methods to fit the multivariate data. 
\begin{enumerate}
\renewcommand{\labelenumi}{(\theenumi)}
\item $ARMA(1,1)-GRACH(1,1)$ \cite{bollerslev1986generalized}, where only $x_{t-1}$ and $\sigma_{t-1}^2$ are included;
\item Gaussian process volatility model (GP-Vol) \cite{Yue2014Nips}, where the Gaussian process for generating $x_t$ is parameterized by $\mu = 0$ and $\sigma^2$;
\item Convolutional neural network (CNN) \cite{schmidhuber2015deep} with three-layered MLPs;
\item LSTM \cite{hochreiter1997long}, which consists of 100 LSTM units and an additional dense layer with 10 neurons;
\item Variational LSTM, i.e., the VRNN in \cite{chung2015recurrent}, where an auto-regressive prior is implemented on latent variables and the RNN part shares the same settings as the above general LSTM, thus we label it  VLSTM.
\end{enumerate}

Here, the comparison between VLSTM and WPVC-VLSTM also serves the purpose of an ablation study.

\subsection{Evaluation Metrics}

Both technical significance and business impact are evaluated. The technical metrics consist of an efficiency evaluation using the relative absolute error (RAE) and the root relative squared error (RSE), and the accuracy assessment including the mean absolute percentage error (MAPE), and the confusion matrix related measures precision, recall, and accuracy.
\begin{enumerate}
\renewcommand{\labelenumi}{(\theenumi)}
\item \textit{RAE} and \textit{RRSE} are both used for analyzing the prediction efficiency in a scaled version.
%      \begin{equation}
%      \begin{split}
%      & RAE = \frac{\sum_{s,t}|\hat{x}_{t,s} - x_{t,s}|}{\sum_{s,t}|mean(x) - x_{t,s}|} \\
%      & RSE = \frac{\sqrt{\sum_{s,t}(\hat{x}_{t,s} - x_{t,s})^2}}{\sqrt{\sum_{s,t}(mean(x) - x_{t,s})^2}} \\
%      \end{split}
%      \end{equation}
\item \textit{MAPE} measures the ratio of the difference between the average absolute error percentage for each time period and the predicted values over the actual values. 
%      \begin{equation}
%      MAPE = \frac{1}{T}\sum|\frac{\hat{x}_{t,s} - x_{t,s}}{x_{t,s}}|
%      \end{equation}
\item Confusion matrix measures the prediction accuracy in terms of \textit{accuracy}, \textit{precision}, and \textit{recall}.
%$TP$, $FP$, $TN$ and $FN$ represent the true positive, false positive, true negative, and false negative, respectively, their derived measures  include:
%      \begin{equation}
%      \begin{split}
%      & Precision = \frac{TP}{TP+FP} \\
%      & Recall = \frac{TP}{TP+FN} \\
%      & Accuracy = \frac{TN+TP}{TP+FP+FN+TN} \\
%      \end {split}
%      \end{equation}
\end{enumerate}

For the confusion matrix, as the measures share duality, all  predictions are transferred to buy and sell actions as described in our business scenarios. We indicate the daily returns with an upward trend as positive in our experiment. The portfolio level assessment is applied  to evaluate the modeling performance of our model. The return $\gamma_{p,t}$ of our portfolio $p$ at time $t$ is calculated by:
\begin{equation}
   \gamma_{p,t} = \sum_{i = 1} ^N \mu_i \gamma_{i,t}
\end{equation}
$\gamma_{i,t}$ is the return of asset $i$ ($i = 1, \ldots, N$) at time $t$, and $\mu_i$ is the holding weight of the asset, where the sum of the holding weights equals 1. To simplify the business scenario, we assume $\mu_i = \frac{1}{N}$, which means each element is weighted the same. 

For portfolio and risk management, we always need to analyze the performance of a portfolio gained by implementing the predictive outcome of each approach applied in trading and the associated risk we need to take. In this experiment, we treat  portfolio $p$ using a generic strategy as follows: at time $t$, when an upward trend is detected according to our prediction, a buy action is taken; otherwise if the trend is predicted as downward, a sell action is taken instead. Two popular metrics from business, the annualized rate of return (ARR), and the value at risk (VaR) measure the business impact of applying this trading strategy $p$:
\begin{enumerate}
\renewcommand{\labelenumi}{(\theenumi)}
\item \textit{ARR} evaluates the performance of portfolio returns over a period scaled down to year-measured trading period, which is defined as:
\begin{equation}
    ARR = \frac{\sum_{i = 1} ^ N \gamma_{p,t}}{T}
\end{equation}
where $T$ denotes the number of trading periods in the selected years, $\gamma_{p,t}$ is the return of portfolio $p$ in the period $t$.
\item \textit{VaR} is an estimate of loss for portfolio investment during a given trading period at a relative confidence level. Given the confidence level $(1 - \alpha)$, the VaR at time $t$ is: 
\begin{equation}
\small
VaR_{1 - \alpha}(\gamma_{p,t}) = -inf\{\gamma | \mathbb{P}(\gamma_{p,t} \leq \gamma_0 | I)  \geq (1 - \alpha)\}
\end{equation}
where $\mathbb{P}$ is the distribution of portfolio return $\gamma_{p,t}$ on level $(1 - \alpha)$, $I$ denotes the past information as a reference. A better risk management strategy intuitively expects a higher VaR.
\end{enumerate}

With the profit or loss of daily return over a given period and the VaR at time $t$, an exceedance function can be defined to determine backtesting:
\begin{equation}
\label{EQ_Exceedance}
    {I_{t+1}} = \left\{ \begin{array}{cc}
    1, & \textrm{if} \quad{\rm{ }}{\gamma_{p,t+1}} < VaR_{(1 - \alpha)} \\
    0, & \textrm{if} \quad{\rm{ }}{\gamma_{p,t+1}} \geqslant VaR_{(1 - \alpha)}
\end{array} \right.
\end{equation}
This sequence of exceedance function can assist us in applying the null hypothesis backtesting framework \cite{Christoffersen1998}  to determine the accuracy of VaR prediction. The backtesting comprises two steps: the unconditional coverage test, and the independence test. First, the unconditional coverage test under the restriction of the occurrence of VaR violations evaluates the loss in excess of the predicted  VaR of the deployed portfolio per Eqn. (\ref{EQ_Exceedance}). Then, the independence test evaluates the aggregation of clusters when these exceedances occur. Finally, the summation of the unconditional coverage test and independence test compose the conditional coverage test, which comprehensively shows the goodness of the VaR model for prediction. The log-likelihood ratios of unconditional coverage (UC) and independence test (IT) are defined as:
\begin{equation}
\label{CB2T_EqDefLRUC}
\begin{split}
& LR_{UC} = -2 log \left( \frac {L(q; I_1,..., I_{T}) }{ L (\hat{\pi}; I_1,..., I_{T}) }   \right) \\
& LR_{IT} = -2 log \left(   \frac{L( \hat{\pi}_2 ; I_1,..., I_{T}) } {   L( \hat{\pi}_{01}, \hat{\pi}_{11} ; I_1,..., I_{T} )} \right) 
%= &-2 log \left(  \frac { \hat{\pi}_2^{n_{01}+n_{11}} (1-\hat{\pi}_2)^{n_{00}+n_{01}} } { \hat{\pi}_{01}^{n_{01}} (1- \hat{\pi}_{01})^{n_{00}}  \hat{\pi}^{n_{11}} ( 1- \hat{\pi}_{11} )^{n_{10}}   }   \right)  \sim \chi_1^2
%\\
%= & -2 log \left(  \frac {q_m(1-q)^{\mathfrak{T}-m}  } { \hat{\pi}^m (1-\hat{\pi})^{\mathfrak{T}-m} }  \right) \sim \chi_1^2 \\
\end{split}
\end{equation}
where $LR_{UC}$ and $LR_{IT}$ denote the log-likelihood ratio of the unconditional coverage and the independence test, respectively. Accordingly, to sum up the result of these two tests, we achieve the log-likelihood ratio of the conditional coverage (CC) test, that is $LR_{CC} = LR_{UC} +LR_{IT}$. Both the unconditional coverage and independence test log-likelihood ratios follow the $\chi^2$ distribution with the degree of freedom equaling 1, while the conditional coverage consequently follows the same distribution but with a two-degree freedom. Accordingly, by having the statistic values of backtesting, a larger $\alpha$ value suggests a more accurate and conscientious result. Given 0.05 as the lower limit of $\alpha$ value, $LR_{UC}$ and $LR_{IT}$ share the critical value 3.841, while the critical value of $LR_{CC}$ is 5.991.

\subsection{Model Specification and Hyperparameter Effect}
We empirically set the hyperparameters of neural networks in our experiments as follows. The LSTM network consists of 100 LSTM units. The neural networks for $g_x$, $g_z$ and $g_y$ for extracting features for inputs, hidden variables and outputs respectively are implemented in a two-layered perceptron (MLP) network with 10 neurons in each layer. The dimension for latent variables is set to 10. The learning rate is  $5 \times 10^{-4}$. The parameters of the model are saved every 10 epochs with the sum over 500 epochs. The raw log returns of daily closing prices are fitted as the input and output of the model. 

\begin{figure}[h]
  \centering
  \tiny
  \subfigure[Log-likelihood]{\label{Trun_Tun:2}\includegraphics[scale = 0.16]{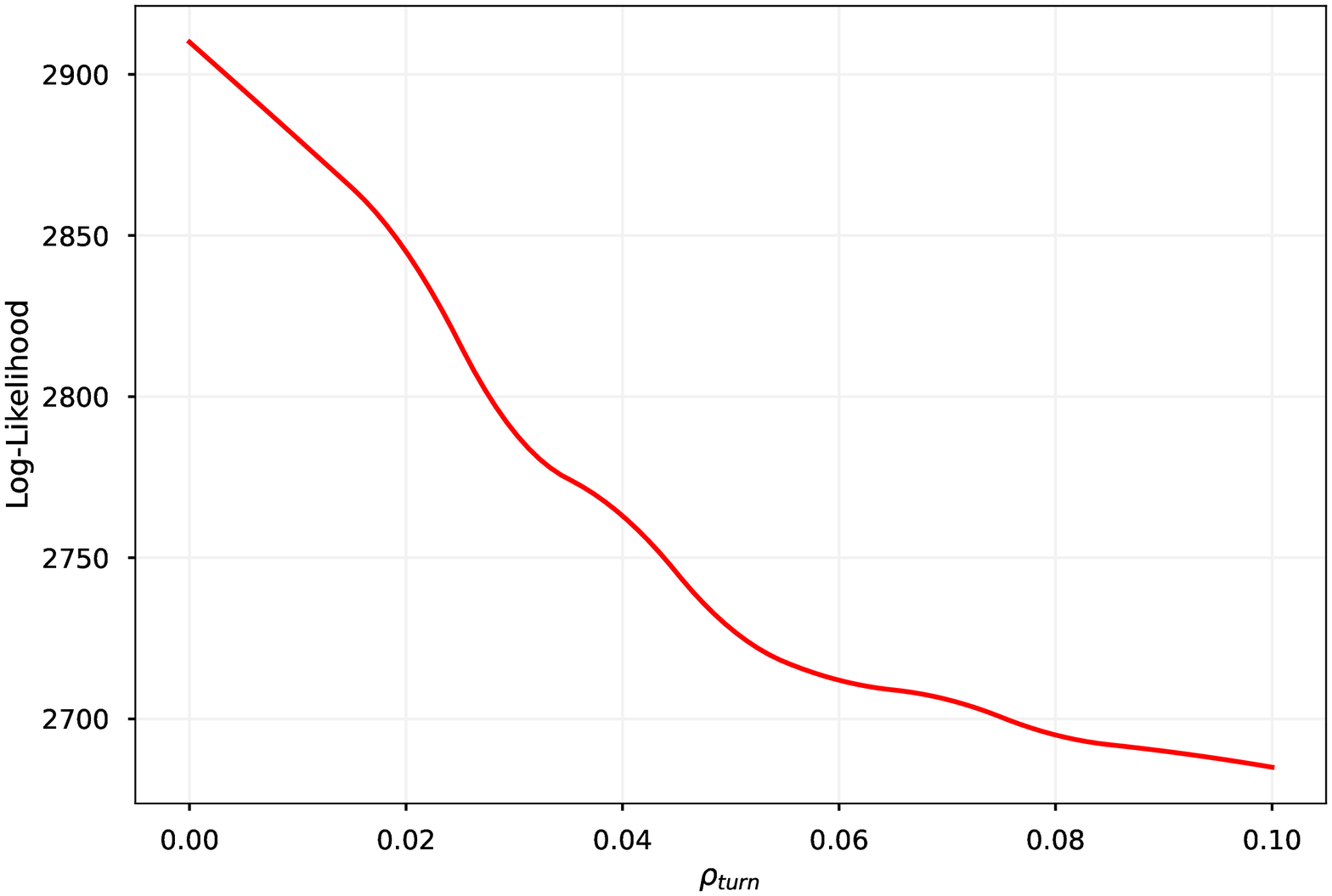}}
  \subfigure[Processing Time]{\label {Trun_Tun:1}\includegraphics[scale = 0.16]{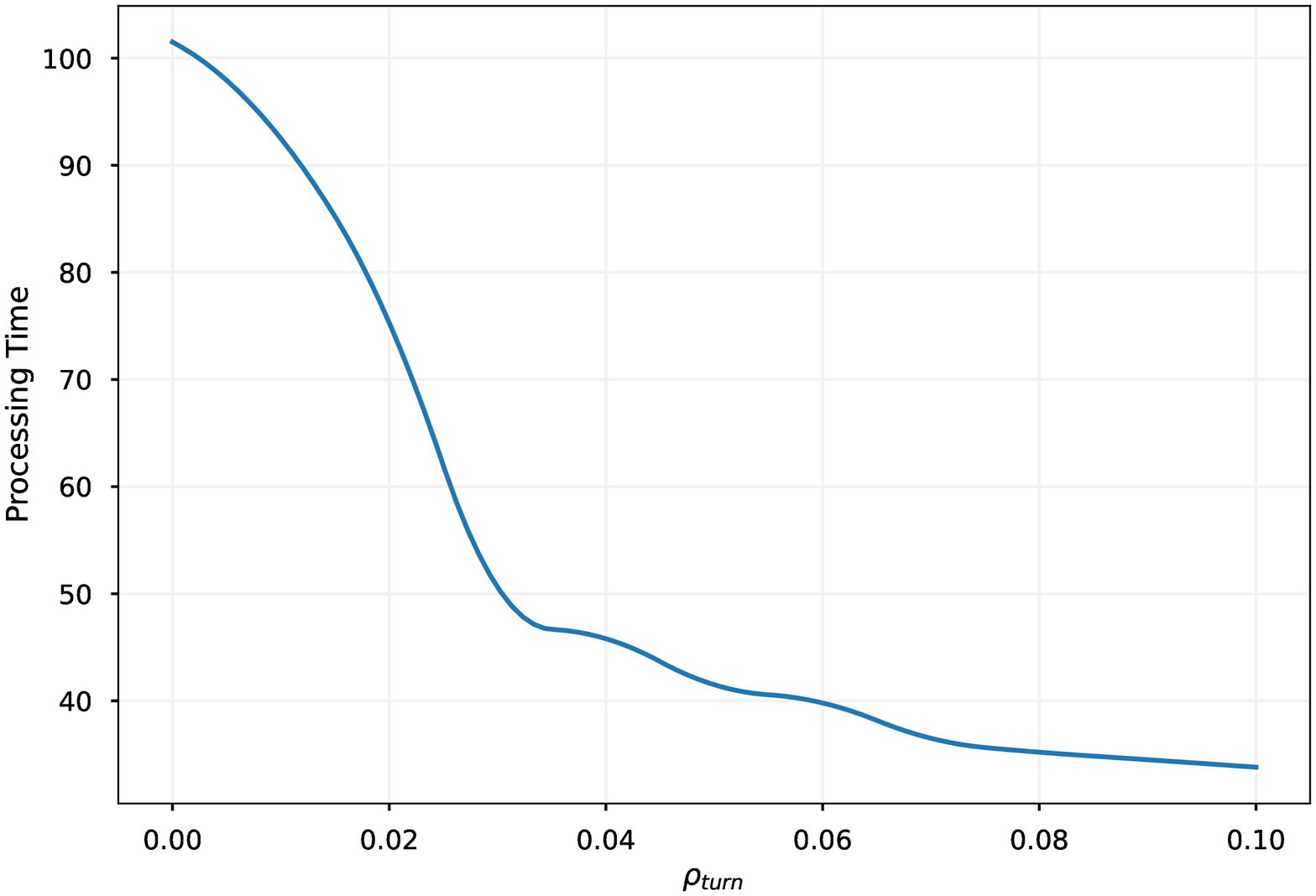}}
  \caption{Effect of the truncation level on the log-likelihood and processing time of WPVC-VLSTM}
  \label{FIG_Trun}
\end{figure}

The regular vine copula involves a hyperparameter truncation rate. We firstly employ the training data to both the non-truncated and truncated models to tune a proper truncation level $\rho_{trun}$. Fig. \ref{FIG_Trun} shows the effect of changing $\rho_{trun}$ on the learning performance log-likelihood and the efficiency in terms of processing time for training the truncated model. The results show a dramatic decrease in the processing time but a flat slop for the log-likelihood value along with the increase of $\rho_{trun}$ at the beginning, then both the processing time and log-likelihood fall gently after $\rho_{trun} = 0.05$. This implies our truncation approach effectively reduces the training cost with little dependence information loss when $\rho_{trun} < 0.05$. Hence, we set the truncation level $\rho_{trun} = 0.05$ in this experiment.

\subsection{Forecasting Performance}
\begin{figure}[t]
\centering
\includegraphics[scale=0.35]{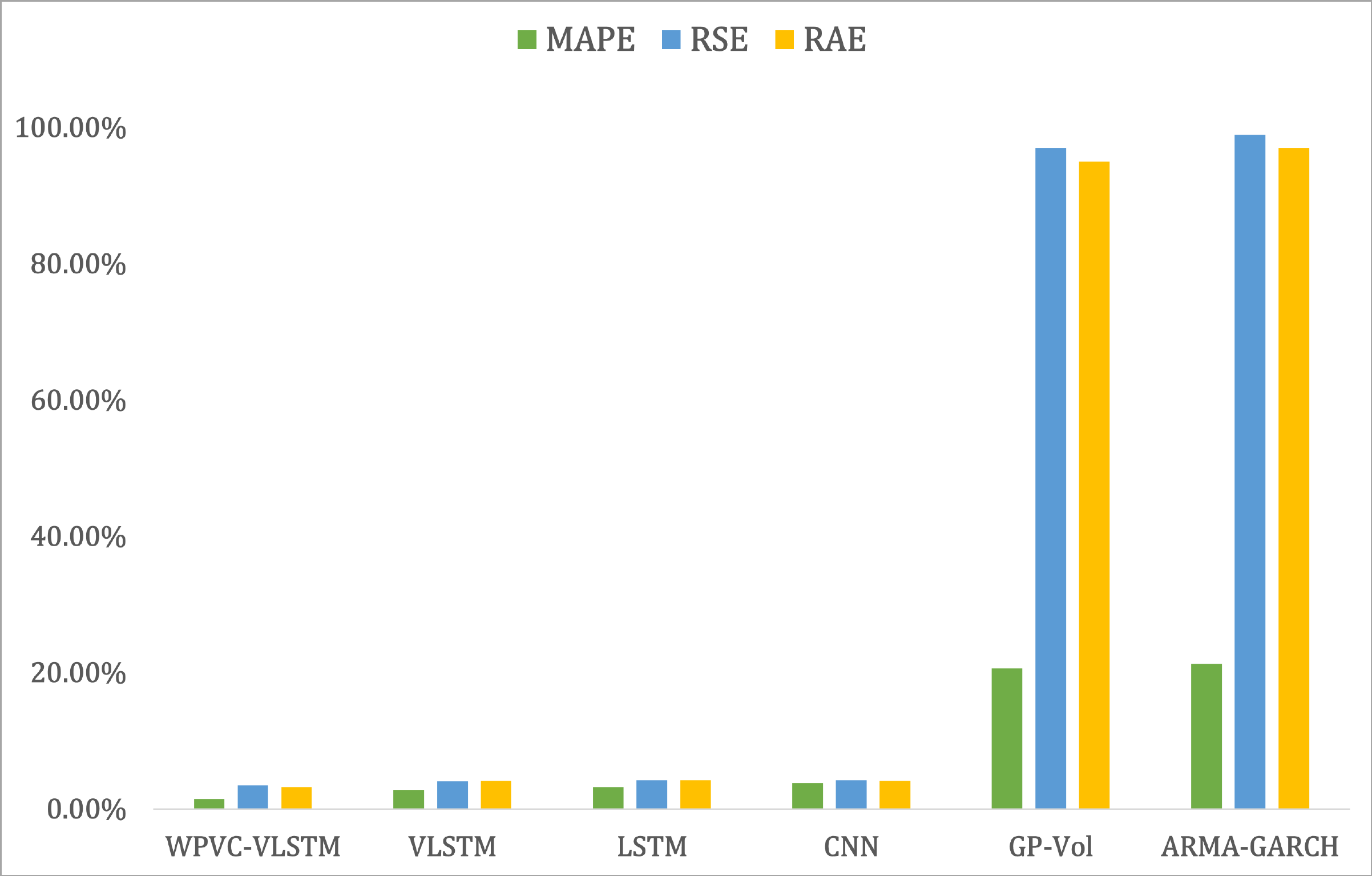}
\caption{The performance evaluation of the portfolio return forecasting}
\label{FIG_Performance}
\end{figure}

We evaluate the forecasting performance of WPVC-VLSTM compared to  ARMA-GRACH, GP-Vol, CNN, LSTM and VLSTM.
Fig. \ref{FIG_Performance} shows the forecasting performance in terms of MAPE, RSE and RAE by comparing the predicted return with the actual one. The results show that WPVC-VLSTM outperforms the other statistical and deep learning approaches. For example, the average MAPE of WPVC-VLSTM only reaches $1.5\%$, which is perceptibly less than ARMA-GARCH and GP-Vol. In addition, WPVC-VLSTM also outperforms the baselines LSTM and CNN by $20.45\%$ and $22.05\%$ in terms of RSE and $30.52\%$ and $32.15\%$ in terms of RAE, respectively. 
%In summary, WPVC-VLSTM achieves better MAPE, RSE and RAE than the baseline models. 

Further, the RAE of WPVC-VLSTM is $15.32\%$ less than the VLSTM with the mean-field assumption, while its RSE is $23.69\%$ smaller. This reveals the effectiveness of the copula variational mechanism. We find that the performance of both WPVC-VLSTM and VLSTM are more stable compared with that of the general LSTM (also confirmed by the results in Tables \ref{TAB_Confusion} and \ref{T_VaR} for more information). This may be due to the introduction of the variational autoencoder mechanism by which the volatility is efficient and the variational networks are more powerful for capturing market uncertainty.

\begin{table}[htbp]
  \centering
  \caption{The performance of ARR and accuracy measures}
    %\begin{tabular}{l|rrrr}
    \scriptsize
    \begin{tabu} to 0.47\textwidth{X[2.0, c] | X[c] X[c] X[c] X[c]}

    \toprule
    Model & \multicolumn{1}{c}{ARR} & \multicolumn{1}{c}{Precision} & \multicolumn{1}{c}{Recall} & \multicolumn{1}{c}{Accuracy} \\
    \hline
    ARMA-GARCH & 12.96\% & 66.74\% & 71.47\% & 66.96\% \\
    GP-Vol      & 20.78\% & 70.67\% & 72.73\% & 69.71\% \\
    LSTM        & 18.52\% & 77.17\% & 79.84\% & 77.35\% \\
    CNN         & 21.56\% & 78.69\% & 82.02\% & 79.71\% \\
    VLSTM       & 22.62\% & 79.73\% & 83.22\% & 80.65\% \\
    WPVC-VLSTM  & \textbf{24.36\%} & \textbf{83.92\%} & \textbf{83.28\%} & \textbf{82.36\%} \\
    \bottomrule
    \end{tabu}
%    \end{tabular}
    \label{TAB_Confusion}
\end{table}

The results of precision, recall, accuracy and ARR are shown in Table \ref{TAB_Confusion}, which illustrate the accuracy-related technical and business performance. The results of precision, recall and accuracy show the discrimination capacity of the models, while ARR validates the achievability of the strategy derived from the predictions on the financial markets when applied in the trading.
The results show that ARMA-GARCH and GP-Vol achieve apparently worse performance compared with other approaches. The main reason for the deficient results is that both approaches are built with the stationary processes, which means the comprehensive hidden dependencies across the financial markets are neglected. Similarly, the general LSTM approach does not perform better than CNN and VLSTM. Note that VLSTM outperforms CNN and LSTM, principally because these two approaches do not involve the volatility with hidden couplings in  financial markets with much volatility. Consequently, WPVC-VLSTM performs better than VLSTM by removing the independent assumptions on the latent variables.

\subsection{Backtesting Performance}

\begin{figure}[htbp]
\centering
\includegraphics[scale = 0.6]{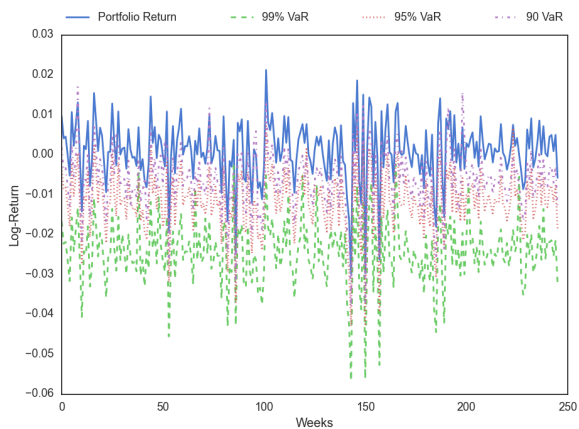}
\caption{The VaR forecasting of portfolio returns per the trading strategies predicted by WPVC-VLSTM}
\label{VaR_Actu}
\end{figure}

The results of backtesting are shown in Table \ref{T_VaR}. Here, POF indicates the prediction failures for each approach, referring to the exceedance determined by strategies in Eqn. (\ref{EQ_Exceedance}). The first row of POF refers to the exact number of exceedances, while the second row provides the percentage of failures. The following sections present the performance of the unconditional coverage $LR_{UC}$, the independence test $LR_{IT}$, and the conditional coverage test $LR_{CC}$, respectively. The first row gives the log-likelihood ratio, while the corresponding confidence level $\alpha$ is provided within the brackets in the second row. Likewise, ARMA-GARCH and GP-Vol achieve the worst performance in the backtesting where the results do not pass the test. The performance of predicted VaR generated by the other four approaches suggests they all effectively represent the volatility hidden in the time series. However, VLSTM and CNN still perform better than LSTM due to embedding the underlying dependencies, and WPVC-VLSTM achieves the best performance over all six models. The predicted return of the portfolio and the associated VaR at different confidence levels by employing WPVC-LTSM are shown in Fig. \ref{VaR_Actu}.

In addition, the results in Tables \ref{TAB_Confusion} and \ref{T_VaR} and Fig. \ref{FIG_Performance} also show the difference between WPVC-VLSTM and VLSTM. They show the effect of incorporating regular vine copula into VLSTM and the effectiveness of integrating probabilistic dependence modeling with variational recurrent neural learning in WPVC-VLSTM for forecasting and backtesting.

\begin{table*}[t]
 \caption{Backtesting results of value at risk ($VaR$) when trading the portfolio}
 \label{T_VaR}
 \centering
 \begin{ThreePartTable}
 \scriptsize
 \begin{tabu} to 0.9\textwidth{X[1.1,c] | X[c] X[c] X[c] X[c] X[c] X[c] X[c]}
 %\begin{tabular}{m{1.8cm}|m{0.7cm}m{0.7cm}m{0.75cm}m{0.75cm}m{0.75cm}}
  \toprule
                                       & $1-\alpha$  & WPVC-VLSTM & VLSTM  & CNN  &  LSTM   & GP-Vol  & ARMA-GARCH \\
  \hline
  \multirow{6}{*}{$POF$}      & \multirow{2}{*}{$99\%$}  & 2         & 4          & 5          & 5         &10         &10                          \\
                                       &                          & 0.81\%    & 1.63\%     & 2.03\%     & 2.03\%    &4.07\%     &4.07\%                \\
                                       \cline{2-8}
                                       & \multirow{2}{*}{$95\%$}  & 12        & 15         & 15         & 17        &35         &40           \\
                                       &                          & 4.88\%    & 6.10\%     & 6.10\%     & 6.91\%    &14.23\%    &16.26\%             \\
                                       \cline{2-8}
                                       & \multirow{2}{*}{$90\%$}  & 23        & 29         & 29         & 31        &61        &59          \\
                                       &                          & 9.35\%    & 11.79\%    & 11.79\%    & 12.60\%   &24.80\%   &23.98\%               \\
  \hline
  \multirow{6}{*}{$LR_{UC}$}  & \multirow{2}{*}{$99\%$}  & 0.032     & 2.186      & 3.376      & 4.770     &10.991      &11.172           \\
                                       &                          & (0.857)   & (0.139)    & (0.066)    & (0.029)   &(0.004)    &(0.004)           \\
                                       \cline{2-8}
                                       & \multirow{2}{*}{$95\%$}  & 0.382     & 0.451      & 0.451      & 0.662      &10.192      &11.735         \\
                                       &                          & (0.536)   &(0.502)     &(0.502)     & (0.416)    &(0.005)    &(0.003)            \\
                                       \cline{2-8}
                                       & \multirow{2}{*}{$90\%$}  & 1.100     & 1.363      & 1.363      & 1.782      &10.456      &11.962                 \\
                                       &                          & (0.294)   & (0.245)    & (0.245)    & (0.182)    &(0.005)    &(0.003)       \\
  \hline
  \multirow{6}{*}{$LR_{IT}$}  & \multirow{2}{*}{$99\%$}  & 2.315     & 2.221      & 1.843      & 1.439      &5.681      &3.980          \\
                                       &                          & (0.128)  & (0.136)    & (0.175)    & (0.230)     &(0.017)    &(0.046)            \\
                                       \cline{2-8}
                                       & \multirow{2}{*}{$95\%$}  & 0.188     & 0.133     & 0.133      & 0.042       &6.032      &4.642         \\
                                       &                          & (0.665)  & (0.715)    &(0.715)     & (0.838)     &(0.014)    &(0.031)      \\
                                       \cline{2-8}
                                       & \multirow{2}{*}{$90\%$}  & 1.582     & 3.533     & 3.533      & 4.469       &6.325      &6.037         \\
                                       &                          & (0.208)   & (0.060)   & (0.060)    & (0.035)     &(0.012)    &(0.014)      \\
  \hline
  \multirow{6}{*}{$LR_{CC}$}  & \multirow{2}{*}{$99\%$}  & 2.347     & 4.408     & 5.218      & 6.209       &16.672     &15.152          \\
                                       &                          & (0.309)   & (0.110)   & (0.074)    & (0.045)     &(0.000)    &(0.001)         \\
                                       \cline{2-8}
                                       & \multirow{2}{*}{$95\%$}  & 0.570     & 0.584     & 0.584      & 0.704       &16.224     &16.377         \\
                                       &                          & (0.752)   & (0.747)   & (0.747)    & (0.703)     &(0.000)    &(0.000)         \\
                                       \cline{2-8}
                                       & \multirow{2}{*}{$90\%$}  & 2.683     & 4.896     & 4.896      & 6.251       &16.781     &17.999         \\
                                       &                          & (0.261)   & (0.086)   & (0.086)    &(0.044)      &(0.000)    &(0.000)         \\
 
  \bottomrule
  %\end{tabular}
  \end{tabu}

  %\begin{tablenotes}
  %    \small
  %    \item[1] Here, PoF is the percentage of exceedance failures. The first row shows the exceeding number, and the second row gives the corresponding percentage;
  %    \item[2] The first row in each cell shows the statistic value, while the corresponding $p$-value is given in the corresponding parenthesis.
  %\end{tablenotes}
  \end{ThreePartTable}
\end{table*}

\section{Discussion}

High-dimensional cross-multivariate dependence modeling is a problem widely applicable to any scenarios and applications where multivariates exist and interact. Cross-market modeling is a typical and challenging example of high-dimensional cross-multivariate dependence modeling, which has been rarely studied in machine learning. CMM is applicable to financial area and any areas where multiple markets, organizations or systems are involved. Examples include the analysis of capital markets, energy markets, utility markets, industry sectors, manufacturing markets, and even tourism markets. Such cross-market problems more or less involve the various challenges discussed in Section \ref{subsec:challenges} in addition to the complexities generic or specific to their underlying domains. 

First, there are other challenges in cross-market settings, such as time-evolving multi-party interactions between non-normal heterogeneous financial markets (variables) and the interaction dynamics. Second, the real-life financial variables present complex data characteristics beyond the non-normal conditions discussed in this paper, such as significant but emergent changes in multivariate time series, inconsistencies across data distributions, temporal and numerical granularities, and the evolving nature of  related systems. Third, modeling the influence of multi-aspect contextual factors such as cultural and geopolitical factors across countries and regions on underlying systems such as equity markets is essential yet challenging. These challenges require more advanced and specific designs for CMM and for high-dimensional cross-multivariate dependence modeling. 

From a modeling and learning perspective, high-dimensional cross-multivariate dependence modeling poses significant theoretical and practical challenges to existing literature. The conventional focus on multivariate time series is constrained by their limited capacity in capturing complex cross-multivariate observable and latent couplings and non-standard dynamics. The recent progress on deep state, factor, relation modeling shows promising in capturing deep features and relations through deep learning, which, however, lacks capacity in customizing data complexities across multivariates. One direction would be to seamlessly integrate statistical relation models such nonparametric Bayesian and time series models such as MGARCH with VAE, LSTM, Transformer or deep sequential models to capture the inconsistencies, heterogeneities and couplings in deep variational frameworks beyond the existing practices.

\section{Concluding Remarks}
We address an important, challenging yet rarely studied problem - high-dimensional cross-multivariate dependence modeling. As a typical example, cross-market modeling aims to capture couplings, co-movement, and co-influence across financial markets and across capital, Internet, and mobile-based economic markets. The modeling of regionalized to globalized multi-market movements must  incorporate and characterize the cross-market couplings for individual or multi-market analysis, asset and portfolio management, and risk management. The recent research shows that conventional multivariate time series and deep neural networks do not essentially capture the explicit and implicit intra- and inter-market couplings. It is even more challenging to model the hierarchical and heterogeneous cross-market interactions in real-life market conditions such as with asymmetric, nonstationary and non-IID market characteristics in sophisticated contexts. 

This work takes a step forward in high-dimensional cross-multivariate dependence modeling. We address the aforementioned cross-market characteristics by characterizing and learning (1) both deep and probabilistic intra- and inter-market couplings using recurrent LSTM empowered by a variational inference-driven  authoencoder, and (2) the asymmetric dependencies between latent variables in the variational inference by a weighted partial vine copula. Our work is the first attempt at the variational modeling of dependence structures with non-normal multivariate dynamics and cross-variable interactions. Our copula variational sequential modeling is evaluated in cross-market portfolio forecasting. Further work may model more hierarchical, evolving and heterogeneous cross-market couplings by non-conventional learning systems. Examples are coupling learning between highly divided commodities, digital currency, equity markets, and markets with time-varying multi-distributional dynamics such as by statistical deep models.

% or


%\appendices

%\section{}

\section*{Acknowledgment}
This work is supported in part by Australian Research Council Discovery Grant DP190101079 and Future Fellowship Grant FT190100734.

\ifCLASSOPTIONcaptionsoff
  \newpage
\fi

\bibliographystyle{IEEEtran}
\bibliography{cvlstm}

\begin{IEEEbiography}[{\includegraphics[width=1in,height=1.25in,clip,keepaspectratio]{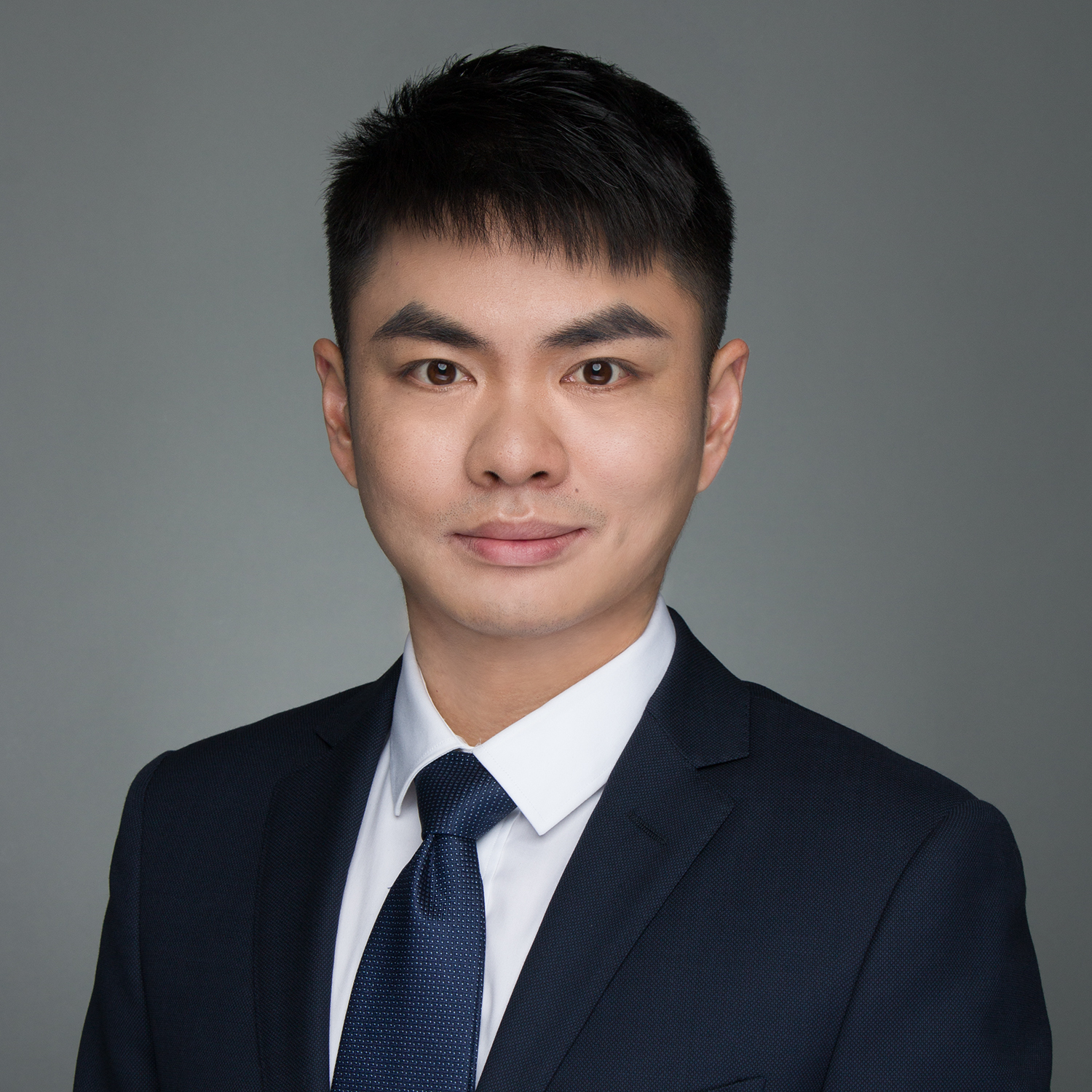}}]{Jia Xu} received a PhD degree in Analytics and works as the head of risk management at a start-up bank. His research mainly focuses on data mining and machine learning and their applications in financial market forecasting and risk management.
\end{IEEEbiography}

\begin{IEEEbiography}[{\includegraphics[width=1in,height=1.25in,clip,keepaspectratio]{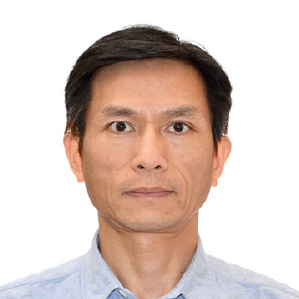}}]{Longbing Cao}(SM'06) received a PhD degree in pattern recognition and intelligent systems and another PhD in computing sciences. He is a Professor at the University of Technology Sydney and an Australian Research Council Future Fellow (Level 3). His  research interests include data science, data mining, machine learning, artificial intelligence and intelligent systems, behavior informatics, and their enterprise applications.
\end{IEEEbiography}


%\vfill


\end{document}